\documentclass[pra,reprint,amsmath,amssymb,aps,longbibliography,superscriptaddress]{revtex4-2}

\usepackage{xcolor}
\usepackage{graphicx}
\usepackage{hyperref}
\usepackage{physics}
\usepackage{siunitx}
\usepackage{ulem}

\usepackage{dcolumn}
\usepackage{graphicx}
\usepackage{mdwlist}
\usepackage{subfigure}
\usepackage{booktabs}
\usepackage{amsmath}
\usepackage{extarrows}
 \usepackage{multirow}
\usepackage{dsfont}
\usepackage{amstext}
\usepackage{amsbsy}
\usepackage{bbm}
\usepackage{bm}
\usepackage{threeparttable}
\usepackage{booktabs}
\usepackage{bbding,pifont}
\usepackage{amsthm}
\usepackage{graphicx}
\usepackage{textcomp}
\usepackage{multirow}
\usepackage{pifont}
\usepackage{color}
\usepackage{sansmath}
\usepackage{makecell}
\usepackage{tabularx}

\hypersetup{
	colorlinks=true,
	urlcolor=purple,
	citecolor=blue,
	linkcolor=purple,
	breaklinks,
    }

\begin{document}

\title{Heralded Generation of Multipartite Free-Electron W-State Entanglement}
\date{\today}

\author{Du Ran}
\email{randu11111@163.com}
\affiliation{School of Electronic Information Engineering, Yangtze Normal University, Chongqing 408100, China\\}
\affiliation{Key Laboratory of Optical Chip and Intelligent Optoelectronic Systems, Chongqing Municipal Education Commission, Yangtze Normal University,  Chongqing 408100, China\\}

\author{Jing-Yi Liu}
\affiliation{School of Electronic Information Engineering, Yangtze Normal University, Chongqing 408100, China\\}

\author{Dan Jin}
\affiliation{School of Electronic Information Engineering, Yangtze Normal University, Chongqing 408100, China\\}

\author{Shuai Liu}
\affiliation{School of Physics and Electronic Engineering, Hubei University Of Arts And Science, Xiangyang, 441053, China\\}

\author{Reuven Ianconescu}
\affiliation{Department of Electrical Engineering Physical Electronics, Tel Aviv University, Ramat Aviv 69978, Israel\\}
\affiliation{Shenkar College of Engineering and Design 12, Anna Frank St., Ramat Gan, Israel\\}

\author{Ze-Long He}
\affiliation{School of Electronic Information Engineering, Yangtze Normal University, Chongqing 408100, China\\}
\affiliation{Key Laboratory of Optical Chip and Intelligent Optoelectronic Systems, Chongqing Municipal Education Commission, Yangtze Normal University,  Chongqing 408100, China\\}

\author{Ji-Yuan Bai}
\affiliation{School of Electronic Information Engineering, Yangtze Normal University, Chongqing 408100, China\\}

\author{Ya-dong Li}
\affiliation{School of Electronic Information Engineering, Yangtze Normal University, Chongqing 408100, China\\}

\author{Avraham Gover}
\email{gover@eng.tau.ac.il}
\affiliation{Department of Electrical Engineering Physical Electronics, Tel Aviv University, Ramat Aviv 69978, Israel\\}

\begin{abstract}
We propose a heralded protocol for generating multipartite free-electron entanglement from atomic $W_N$ resources in a sideband-resolved interaction regime.
The scheme consists of $N$ independent electron--atom interaction arms, where each free electron couples locally to one two-level system.
For uniform couplings and common detuning, the dynamics is solved analytically within the rotating-wave approximation.
Projecting the atoms onto the all-ground state maps the initial atomic excitation manifold onto the electronic upper-sideband manifold and prepares an exact $N$-electron $W_N$-type state.
The heralding probability is obtained in closed form for resonant and detuned regimes.
At resonance, the optimal success probability obeys the large-$N$ scaling $P_{G_N}^{\max}\sim e^{-1}/N$.
The heralded state retains the multipartite entanglement structure of the atomic resource, as shown for arbitrary $N$ and illustrated explicitly for $N=3$.
Detuning, weak symmetry breaking, beyond-rotating-wave corrections, and Gaussian coupling envelopes are discussed.
The protocol provides a scalable route toward multipartite free-electron entanglement generation from localized atomic resources within quantum electron optics.
\end{abstract}

\maketitle

\section{Introduction}

Quantum entanglement is a central resource for quantum teleportation \cite{luo2019quantum}, quantum cryptography \cite{ekert1991quantum}, quantum communication \cite{bennett1993teleporting}, quantum computing \cite{gottesman1999demonstrating}, and tests of quantum mechanics \cite{aspect1982experimental,paneru2021experimental}.
Over the past decades, entangled states have been generated in a broad range of physical platforms, including photons \cite{forbes2025heralded,chen2021quantum},  neutral atoms \cite{wilk2010entanglement,shao2017ground}, superconducting circuits \cite{chen2025hardware,zhang2023generating}, and cavity quantum electrodynamics (cavity QED) systems  \cite{zheng2000efficient,zheng2001one}.
Experimental and theoretical progress has enabled the preparation of Bell states \cite{yuan2020steady,zou2022bell,weng2025high}, Greenberger--Horne--Zeilinger states \cite{yang2018deterministic,zhang2024fast}, and $W$ states \cite{peng2021one,zang2016deterministic} through a variety of mechanisms, including quantum Zeno dynamics \cite{barontini2015deterministic,liang2015adiabatic,wang2008quantum}, adiabatic passage \cite{chang2020remote,carrasco2024dicke,xu2024efficient}, measurement-feedback control \cite{francesco2024steady,hashim2025efficient}, and dissipation engineering \cite{kastoryano2011dissipative,cole2022resource,zhang2023generation}.
Among available routes to entangled-state preparation, entanglement transfer is especially useful because controlled interactions can map quantum correlations from a prepared resource to a different physical carrier \cite{paternostro2004complete,lee2006entanglement,serafini2006enhanced,zou2006entanglement,casagrande2007improving}.
A central obstacle to selective atom--field entanglement transfer at individual targets is the limited spatial resolution of optical addressing, which remains challenging for implementations requiring single-TLS control.

More recently, quantum electron optics (QEO) has emerged as a platform for coherent preparation and manipulation of free-electron wave packets through photon-induced near-field electron microscopy (PINEM), electron energy-loss spectroscopy (EELS), and cathodoluminescence (CL) \cite{reinhardt2020theory,zhang2025spontaneous,madan2022ultrafast,vanacore2018attosecond,wong2021control,vanacore2020spatio}.
QEO enables ultrafast nanoscale spectroscopy and quantum-state engineering by exploiting shaped electron--matter interactions to imprint, probe, and herald quantum correlations in emitted or scattered bosonic fields \cite{di2019probing,kfir2021optical,sun2023generating,ben2021shaping,tsarev2021measurement,garcia2021optical,arend2025electrons}.
Direct experimental verification of free-electron--photon entanglement has been achieved through coincidence-based measurements and quantum imaging protocols in transmission electron microscopy platforms \cite{henke2025observation,preimesberger2025experimental}.
These advances establish QEO as a realistic platform for generating, manipulating, and probing quantum correlations involving free-electron wave packets.

In particular, free-electron--bound-electron resonant interaction (FEBERI) provides a well-defined mechanism for coupling a quantum electron wave packet (QEW) to an atomic two-level system (TLS), with interaction strength governed by the longitudinal density modulation of the QEW \cite{gover2020free}.
Building on such a mechanism, engineered QEWs enable efficient excitation and quantum-state interrogation of TLSs \cite{zhang2021quantum,zhang2022quantum,ran2022coherent}, while recent studies have further extended free-electron quantum interactions to a broader range of bosonic and material excitations \cite{zhao2021quantum,ruimy2021toward,crispin2025probing,yalunin2021tailored,morimoto2021coherent,ratzel2021controlling}.
While recent studies have mainly focused on quantum correlations and entanglement between free electrons and photonic degrees of freedom, the controlled generation of multipartite entanglement directly between free electrons remains largely unexplored.
In parallel, entanglement-swapping and quantum-eraser protocols have also been proposed as promising approaches for exploiting electron-mediated quantum correlations in nanoscale probing and quantum-enhanced measurements \cite{henke2025probing}.
Free-electron entanglement is of particular interest because entangled electron pairs can support superradiant and subradiant light emission \cite{karnieli2021superradiance}, while symmetric multi-electron entangled states can enable enhanced deterministic photon generation \cite{karnieli2023jaynes}.

Building on the recent demonstration of heralded entanglement transfer from an entangled atomic pair to two free electrons \cite{ran2026heralded}, we generalize the protocol to the multipartite setting.
In particular, we investigate the heralded transfer of multipartite entanglement from an initially entangled ensemble of two-level systems to free electrons, with the aim of generating multipartite entangled free-electron states.
We consider an N-arm local-interaction geometry in which each electron couples only to its corresponding TLS, while the atomic subsystem is initially prepared in the single-excitation $W_N$ state.
Under symmetric conditions,  the multipartite dynamics is derived analytically in the rotating-wave approximation (RWA), and postselection of the all-ground TLS outcome prepares a heralded multipartite entangled free-electron state in the upper-sideband single-excitation manifold.
The transfer properties are analyzed for general $N$, while the tripartite case $N=3$ illustrates the resulting entanglement structure and correlation redistribution explicitly.
The effects of detuning, weak symmetry breaking, corrections beyond the RWA, and Gaussian coupling envelopes are also examined.
The present work establishes a concrete route toward heralded multipartite free-electron entanglement generation and extends entanglement-transfer protocols into the emerging framework of QEO.

The remainder of the paper is organized as follows.
In Section~\ref{SystM}, we introduce the system model and derive the single-arm detuned rotating-wave dynamics.
Section~\ref{HET} develops the general framework of multipartite heralded entanglement transfer and includes the success probability, the entanglement structure of the heralded state, and the explicit tripartite example.
The physical implications of the protocol, together with its robustness and limitations, are discussed in Section~\ref{DISS}.
Section~\ref{CON} concludes the paper.

\section{System Model}\label{SystM}

We consider $N$ free electrons interacting with $N$ nearby two-level systems (TLSs), with $j=1,2,\dots,N$ labeling each electron--TLS pair.
Each free electron couples only to its corresponding TLS, forming a local interaction geometry with $N$ independent arms.
The electrons are treated as distinguishable wave packets associated with different spatial arms, and direct inter-arm couplings are neglected.
Such a spatially local interaction picture is natural in free-electron quantum-optical settings, where an electron couples to a localized quantum system through a near field or an effective dipolar interaction \cite{gover2020free,zhang2022quantum}.
The resulting dynamics can therefore be built from elementary local electron--TLS exchange processes.
In the sideband-resolved description, the natural basis for the $j$th electron is the discrete ladder $\{|n\rangle_j\}_{n\in\mathbb{Z}}$, where neighboring basis states differ by one exchanged quantum $\hbar\omega$.
The corresponding sideband-number operator is \(\hat n_j=\sum_{n=-\infty}^{\infty}n|n\rangle_j\,{}_j\langle n|\).
Such a representation is standard in free-electron quantum optics whenever the interaction induces coherent single-quantum transitions between adjacent electron energy sidebands \cite{bucher2023free,tsesses2023tunable}.
We therefore define the sideband translation operators $\hat b_j  = \sum_{n=-\infty}^{\infty}|n-1\rangle_j\,{}_j\langle n|$,
and
$\hat b_j^\dagger  = \sum_{n=-\infty}^{\infty}|n+1\rangle_j\,{}_j\langle n|$, which satisfy $\hat b_j|n\rangle_j = |n-1\rangle_j$, $\hat b_j^\dagger|n\rangle_j = |n+1\rangle_j$ and $\hat b_j^\dagger\hat b_j = \hat b_j\hat b_j^\dagger=\hat I$.
The operators $\hat b_j$ and $\hat b_j^\dagger$ therefore translate the electron along the sideband ladder by one quantum of energy.

For the $j$th TLS, we denote the ground and excited states by $|g\rangle_j$ and $|e\rangle_j$, and define the Pauli ladder operators $\hat\sigma_j^+ = |e\rangle_j\langle g|$, $\hat\sigma_j^- = |g\rangle_j\langle e|$, and $\hat\sigma_j^z = |e\rangle_j\langle e|-|g\rangle_j\langle g|$.
The free Hamiltonian of the considered system is
\begin{equation}
\hat H_0=\sum_{j=1}^{N}\hbar\omega\hat n_j+\sum_{j=1}^{N}\frac{\hbar\omega_0}{2}\hat\sigma_j^z,
\label{eq:H0}
\end{equation}
where $\omega_0$ is the TLS transition frequency.
The first term describes the free evolution of the electron sideband ladder, while the second term gives the bare TLS splitting.
The bilinear interaction Hamiltonian reads
\begin{equation}
\hat H_{\mathrm{int}}(t)=\sum_{j=1}^{N}\hbar G_j(t)\bigl(\hat b_j+\hat b_j^\dagger\bigr)\bigl(\hat\sigma_j^++\hat\sigma_j^-\bigr),
\label{eq:Hint}
\end{equation}
where $G_j(t)$ denotes the effective time-dependent coupling strength in the $j$th local interaction arm, determined by the local electron--TLS interaction geometry and by the finite interaction window during which the electron passes the corresponding TLS.
Equation~(\ref{eq:Hint}) then represents single-quantum exchange between neighboring electron sidebands and the two-level transition of the local TLS.
In the interaction picture with respect to $\hat H_0$, the interaction Hamiltonian becomes
\begin{equation}
\begin{split}
\hat H_I(t)=\sum_{j=1}^{N}\hbar G_j(t)\Bigl[
\hat b_j\hat\sigma_j^+e^{-i\Delta t}
+\hat b_j^\dagger\hat\sigma_j^-e^{+i\Delta t}
+\hat b_j\hat\sigma_j^-e^{-i\Omega_+ t}\\
+\hat b_j^\dagger\hat\sigma_j^+e^{+i\Omega_+ t}
\Bigr],
\end{split}
\label{eq:HI_full}
\end{equation}
where $\Delta = \omega-\omega_0$ and $\Omega_+ = \omega+\omega_0$.
The first two terms describe excitation exchange between the electron sideband ladder and the TLS.
The last two terms are counter-rotating contributions oscillating at the fast frequency $\Omega_+$.
In the near-resonant regime $|\Delta|\ll \omega,\omega_0$, the rotating-wave approximation (RWA) gives
\begin{equation}
\hat H_{I,\mathrm{RWA}}(t)=\sum_{j=1}^{N}\hbar G_j(t)\Bigl[
\hat b_j\hat\sigma_j^+e^{-i\Delta t}
+\hat b_j^\dagger\hat\sigma_j^-e^{+i\Delta t}
\Bigr].
\label{eq:HI_RWA}
\end{equation}
Within this approximation, each local arm supports only the near-resonant single-quantum transfer channel.

We assume that all incoming electrons are prepared in the same reference sideband $|n_0\rangle$, namely $|\psi_E(0)\rangle=|n_0\rangle_1|n_0\rangle_2\cdots|n_0\rangle_N$.
For the atomic TLS sector, we consider the general single-excitation $W_N$-type resource
\begin{equation}
|W_N(\boldsymbol{\phi})\rangle_{\mathrm{TLS}}
=\frac{1}{\sqrt N}\sum_{j=1}^{N}e^{i\phi_j}|g\cdots e_j\cdots g\rangle,
\label{eq:WN_tls}
\end{equation}
where $\phi_j$ are relative phases, and the symmetric $W_N$ state corresponds to $\phi_j=0$ for all $j$.
The preparation of $W$-type resources is supported by substantial progress in atomic and atomic-like platforms.
Multipartite $W$-type entanglement has been experimentally generated and fully characterized for four to eight trapped ions \cite{haffner2005scalable}, while deterministic generation of large-scale atomic $W$ states in cavity QED \cite{zang2016deterministic} and dissipative preparation of tripartite atomic $W$ states in Rydberg-atom-cavity systems \cite{li2018dissipation} have also been proposed.
The total initial state of the system is $|\Psi(0)\rangle=|\psi_E(0)\rangle\otimes|W_N(\boldsymbol{\phi})\rangle_{\mathrm{TLS}}$.

In the following, we first consider the symmetric special case $G_j(t) = G(t)$ ($j=1,2,\dots,N$), for which all local interaction arms have identical coupling envelopes.
Because the RWA Hamiltonian is a sum of mutually commuting local-arm generators acting on distinct tensor factors, the full propagator factorizes as $\hat U_{\mathrm{RWA}}(t)=\prod_{j=1}^{N}\hat U_j(t)$, where $\hat U_j(t)$ is the RWA propagator associated with the $j$th local interaction arm.
The dynamics of each arm is therefore determined independently by the corresponding single-arm propagator.
We model the effective local coupling envelope by a ``square pulse",
\begin{align}
G(t) &=
\begin{cases}
G_0, & 0\le t\le T,\\
0, & \mathrm{otherwise}.
\end{cases}
\label{eq:square_pulse}
\end{align}
The accumulated pulse area is \(g(t)=\int_0^t G(t')dt'=G_0t\) inside the interaction window, with \(g\equiv G_0T\).
In a more realistic description, the local coupling envelope is generally smooth and determined by the electron trajectory and the spatial profile of the electron--TLS interaction region \cite{gover2020free,zhao2021quantum}.

For the detuned single-arm dynamics, we introduce the effective detuning parameter and the generalized pulse area
\begin{align}
\delta(t) = \frac{\Delta t}{2}, \ \
\tilde g(t)  = \sqrt{g^2(t)+\delta^2(t)},
\label{eq:delta_tildeg}
\end{align}
as well as the associated local survival and transfer amplitudes
\begin{align}
c_-(t)&=\cos\tilde g(t)-i\frac{\delta(t)}{\tilde g(t)}\sin\tilde g(t), \notag\\
c_+(t)&=\cos\tilde g(t)+i\frac{\delta(t)}{\tilde g(t)}\sin\tilde g(t), \notag\\
s(t)&=\frac{g(t)}{\tilde g(t)}\sin\tilde g(t).
\label{eq:st}
\end{align}

Under the action of the single-arm propagator $\hat U_j(t)$ generated by the $j$th local RWA Hamiltonian, the local basis states in the invariant subspace $\{|n,g\rangle_j,\ |n-1,e\rangle_j\}$ evolve as
\begin{align}
|n,g\rangle_j\longrightarrow c_-(t)|n,g\rangle_j-i\,s(t)|n-1,e\rangle_j,\notag\\
|n-1,e\rangle_j\longrightarrow c_+(t)|n-1,e\rangle_j-i\,s(t)|n,g\rangle_j,
\label{eq:local_nminus1e}
\end{align}
whereas, in the equivalent invariant subspace $\{|n,e\rangle_j,\ |n+1,g\rangle_j\}$, the evolution is
\begin{align}
|n,e\rangle_j\longrightarrow c_+(t)|n,e\rangle_j-i\,s(t)|n+1,g\rangle_j,\notag\\
|n+1,g\rangle_j\longrightarrow c_-(t)|n+1,g\rangle_j-i\,s(t)|n,e\rangle_j.
\label{eq:local_nplus1g}
\end{align}
At exact resonance $\Delta=0$, the detuning-dependent amplitudes introduced in Eq.~(\ref{eq:st}) simplify to
\begin{align}
c_-(t) = c_+(t)=\cos g(t), \ \
s(t)  = \sin g(t).
\label{eq:resonant_coeffs}
\end{align}

For initial conditions in the reference sideband $|n_0\rangle_j$, the local RWA propagator yields
\begin{align}
|n_0,g\rangle_j\longrightarrow c_-(t)|n_0,g\rangle_j-i\,s(t)|n_0-1,e\rangle_j,\notag\\
|n_0,e\rangle_j\longrightarrow c_+(t)|n_0,e\rangle_j-i\,s(t)|n_0+1,g\rangle_j.
\label{eq:n0e_map}
\end{align}
To reconstruct the full multipartite evolution, it is convenient to introduce the $N$-electron basis states
\begin{equation}
|E_j^{(\pm)}\rangle
=
|n_0\rangle_1\cdots|n_0\pm1\rangle_j\cdots|n_0\rangle_N,
\label{eq:Ejplus}
\end{equation}
for which only the $j$th electron is shifted upward ($+$) or downward ($-$) by one sideband.
The set $\{|E_j^{(+)}\rangle\}_{j=1}^N$ spans the upper-sideband single-excitation manifold generated by the local RWA transfer process, whereas the set $\{|E_j^{(-)}\rangle\}_{j=1}^N$ describes the corresponding lower-sideband sector associated with excitation of a TLS by the local electron.

For the initial TLS component $|g\cdots e_j\cdots g\rangle$, the $j$th arm is the only arm that initially carries the atomic excitation.
Consequently, the $j$th arm is the only arm that can transfer that excitation into the upper electron sideband, whereas all remaining arms contribute through local survival amplitudes and lower-sideband excitation channels according to Eq.~(\ref{eq:n0e_map}).
The resulting evolution of the basis component $|n_0\rangle^{\otimes N}\otimes|g\cdots e_j\cdots g\rangle$ takes the form
\begin{equation}
\begin{split}
|n_0\rangle^{\otimes N}\otimes|g\cdots e_j\cdots g\rangle
\longrightarrow
\Bigl[c_+(t)|n_0,e\rangle_j-i\,s(t)|n_0+1,g\rangle_j\Bigr]\\
\times
\prod_{k\neq j}
\Bigl[c_-(t)|n_0,g\rangle_k-i\,s(t)|n_0-1,e\rangle_k\Bigr],
\end{split}
\label{eq:fixed_j_evolution}
\end{equation}
which displays the branch structure of the multipartite dynamics.
For each fixed value of $j$, the upper-sideband electron channel is tied to the unique arm that initially contains the atomic excitation, while all spectator arms contribute through survival amplitudes and lower-sideband leakage channels.
Summing over all $j$ with the phases $e^{i\phi_j}/\sqrt N$ from the initial atomic $W_N$ resource, the total state becomes
\begin{equation}
\begin{split}
|\Psi(t)\rangle
&=
\frac{1}{\sqrt N}
\sum_{j=1}^{N}
e^{i\phi_j}
\Bigl[c_+(t)|n_0,e\rangle_j-i\,s(t)|n_0+1,g\rangle_j\Bigr]\\
&\quad\times
\prod_{k\neq j}
\Bigl[c_-(t)|n_0,g\rangle_k-i\,s(t)|n_0-1,e\rangle_k\Bigr],
\end{split}
\label{eq:psi_t_full}
\end{equation}
which gives the complete multipartite state reconstruction and contains all dynamically generated output sectors.
Such sectors include the all-ground TLS branch, branches with residual TLS excitations, and branches associated with lower-sideband electron population.

\section{Heralded Entanglement Transfer}\label{HET}

\subsection{Heralded $N$-electron branch}

We now distinguish the conditional quality of the heralded electron state from the unconditional efficiency of the full protocol.
The conditional analysis characterizes the electron state associated with the selected measurement outcome, while the unconditional analysis quantifies the probability weight of the selected branch within the full many-body evolution.
Keeping these two quantities separate provides a clear basis for identifying the heralded state and evaluating the corresponding success probability.

Projection of the total state $|\Psi(t)\rangle$ in Eq. (\ref{eq:psi_t_full}) onto the TLS outcome
\begin{equation}
|G_N\rangle_{\mathrm{TLS}}\equiv|g\rangle_1|g\rangle_2\cdots|g\rangle_N,
\label{eq:GN_def}
\end{equation}
selects the branch in which all TLSs end in the ground state after the interaction.
Within the present single-excitation setting, the all-ground TLS outcome imposes a stringent dynamical constraint on the multipartite evolution.
The arm that initially carries the atomic excitation must transfer that excitation to the corresponding electron upper sideband, whereas every remaining arm must stay in the local ground-sideband sector without generating an additional TLS excitation.

According to Eq.~(\ref{eq:psi_t_full}), survival of the all-ground TLS branch requires, for each fixed value of $j$, selection of the transfer amplitude $-i\,s(t)$ from the arm initially prepared in $|e\rangle_j$ and selection of the survival amplitude $c_-(t)$ from every arm initially prepared in $|g\rangle_k$ with $k\neq j$.
All lower-sideband channels are therefore removed by the projection onto $|G_N\rangle_{\mathrm{TLS}}$, because every lower-sideband contribution is accompanied by at least one residual TLS excitation in the final state.
Projection onto the all-ground TLS sector thus isolates the coherent upper-sideband image of the initial atomic single-excitation manifold.

The resulting unnormalized heralded electron state is
\begin{equation}
|\Psi_{G_N}^{(e)}(t)\rangle
=
\frac{-i\,s(t)\,[c_-(t)]^{N-1}}{\sqrt N}
\sum_{j=1}^{N}e^{i\phi_j}|E_j^{(+)}\rangle.
\label{eq:psi_GN_e}
\end{equation}
The factor $-i\,s(t)$ represents the unique local transfer event by which the initial atomic excitation is converted into an upper-sideband electron excitation.
The factor $[c_-(t)]^{N-1}$ represents the joint survival amplitude of the remaining $N-1$ spectator arms, each of which must remain in the local ground-sideband sector in order to contribute to the heralded branch.
The associated heralding probability is
\begin{align}
P_{G_N}(t)
 =
\langle\Psi_{G_N}^{(e)}(t)|\Psi_{G_N}^{(e)}(t)\rangle
 =
|s(t)|^2\,|c_-(t)|^{2N-2}.
\label{eq:PGN_t}
\end{align}
The factor $|s(t)|^2$ gives the probability of the actual excitation-transfer event on the unique initially excited arm.
Each of the remaining $N-1$ arms contributes a factor $|c_-(t)|^2$, corresponding to the requirement that no unwanted local TLS excitation is generated on any spectator arm.
The heralding probability is therefore governed by a single successful transfer event  accompanied by $N-1$ simultaneous local survival conditions.
After normalization, the conditional electron state in Eq. (\ref{eq:psi_GN_e}) becomes
\begin{equation}
|\psi_{G_N}^{(e)}(t)\rangle
=
\frac{1}{\sqrt N}
\sum_{j=1}^{N}e^{i\phi_j}|E_j^{(+)}\rangle.
\label{eq:psi_GN_norm}
\end{equation}
In the symmetric detuned RWA case, Eq.~(\ref{eq:psi_GN_norm}) is exactly an $N$-electron $W_N$-type state in the upper-sideband single-excitation manifold.
The phase factors $\phi_j$ are inherited directly from the initial atomic resource, so the heralded branch preserves both the single-excitation structure and the internal phase coherence of the input state.
The normalized conditional state is independent of time and detuning, whereas the weight of the heralded branch is controlled by $s(t)$ and $c_-(t)$ through Eq.~(\ref{eq:PGN_t}).
For the standard symmetric resource with $\phi_j=0$, Eq.~(\ref{eq:psi_GN_norm}) reduces to
\begin{equation}
|W_N^{(+)}\rangle_e
=
\frac{1}{\sqrt N}
\sum_{j=1}^{N}|E_j^{(+)}\rangle.
\label{eq:WNplus_e}
\end{equation}
In the fully symmetric case, the heralded branch realizes an exact mapping from the atomic single-excitation $W_N$ manifold to the electron upper-sideband single-excitation manifold.

Accordingly, the conditional fidelity of the heralded electron state can be defined as
\begin{equation}
F_{N}^{\mathrm{(cond)}}(t)
=
{}_e\langle W_N^{(+)}|
\hat\rho_{e|G_N}(t)
|W_N^{(+)}\rangle_e,
\label{eq:Fcond_general}
\end{equation}
where $\hat\rho_{e|G_N}(t)
=
\frac{|\Psi_{G_N}^{(e)}(t)\rangle\langle \Psi_{G_N}^{(e)}(t)|}
{P_{G_N}(t)}$.
Within the symmetric detuned RWA treatment, the normalized heralded branch is exactly equal to the ideal target state,
\begin{equation}
\hat\rho_{e|G_N}(t)
=
|W_N^{(+)}\rangle_e\langle W_N^{(+)}|,
\label{eq:rho_cond_ideal_general}
\end{equation}
and therefore
\begin{equation}
F_N^{\mathrm{(cond)}}(t)=1.
\label{eq:Fcond_one_general}
\end{equation}
The unit conditional fidelity confirms that the heralded electron state is pure and that the phase pattern of the initial atomic resource is transferred without distortion into the electron upper-sideband manifold.
The heralded protocol therefore produces a coherent delocalization of a single electron-sideband excitation over all $N$ arms rather than an incoherent mixture of different excitation locations.
The multipartite $W_N$ structure of the heralded electron state is encoded precisely in that coherent superposition.

\subsection{Success probability, optimum, and large-$N$ scaling}

For the square pulse in Eq. (\ref{eq:square_pulse}) evaluated at $t=T$, the dimensionless pulse area, detuning parameter, and generalized pulse area are
\begin{equation}
g  = G_0T,\ \ \delta = \frac{\Delta T}{2},\ \ \tilde g = \sqrt{g^2+\delta^2}.
\label{eq:g_delta_tildeg_T}
\end{equation}
Equation~(\ref{eq:st}) yields
\begin{equation}
|c_-(T)|^2
=
\cos^2\tilde g+\frac{\delta^2}{\tilde g^2}\sin^2\tilde g,\ \
s(T)=\frac{g}{\tilde g}\sin\tilde g .
\label{eq:cminusT_abs2}
\end{equation}
Substitution of Eq.~(\ref{eq:cminusT_abs2}) into Eq.~(\ref{eq:PGN_t}) gives the heralding probability at the end of the interaction window,
\begin{equation}
P_{G_N}(T)
=
\frac{g^2}{\tilde g^2}\sin^2\tilde g
\left(
\cos^2\tilde g+\frac{\delta^2}{\tilde g^2}\sin^2\tilde g
\right)^{N-1}.
\label{eq:PGN_detuned}
\end{equation}
The expression makes explicit two dynamical requirements of the heralded branch.
The arm initially carrying the atomic excitation must provide the local conversion amplitude $s(T)$ that populates the electron upper-sideband channel.
The remaining $N-1$ arms must instead contribute the local survival amplitude $c_-(T)$, excluding residual TLS excitation and lower-sideband leakage.
Thus, the heralding probability is determined by one successful transfer event together with $N-1$ simultaneous spectator-arm survival events.
At exact resonance, namely $\Delta=0$, one has $\delta=0$ and $\tilde g=g$, so Eq.~(\ref{eq:PGN_detuned}) reduces to
\begin{equation}
P_{G_N}(T)=\sin^2 g\,\cos^{2N-2} g.
\label{eq:PGN_resonant}
\end{equation}
The resonant heralding probability is therefore reduced by the increasing number of simultaneous spectator-arm constraints as $N$ grows.

For optimization in the resonant regime, Eq.~(\ref{eq:PGN_resonant}) may be rewritten as $P_{G_N}(g)=\sin^2 g\,\cos^{2N-2} g$.
Introducing the variable $x=\cos^2 g$, one obtains $P_{G_N}(x)=(1-x)x^{N-1}$.
The stationarity condition ${dP_{G_N}}/{dx}=0$ yields $x= (N-1)/{N}$.
The optimal pulse area therefore satisfies
\begin{align}
\cos^2 g_{\mathrm{opt}}  = \frac{N-1}{N}, \  \
\sin^2 g_{\mathrm{opt}}  = \frac{1}{N},
\label{eq:g_opt}
\end{align}
and the maximal heralding probability becomes
\begin{equation}
P_{G_N}^{\max}
=
\frac{1}{N}\left(\frac{N-1}{N}\right)^{N-1}.
\label{eq:PGN_max}
\end{equation}
For large $N$, Eq.~(\ref{eq:PGN_max}) approaches
\begin{equation}
P_{G_N}^{\max}\sim \frac{e^{-1}}{N}.
\label{eq:PGN_asymp}
\end{equation}
The large-$N$ behavior indicates analytical scalability to arbitrary system size, although the optimal success probability decreases algebraically with $N$.
The algebraic scaling has a simple physical origin.
The heralded branch requires only one successful transfer event, accompanied by survival of the remaining $N-1$ spectator arms, rather than $N$ independent transfer events.
Increasing $N$ therefore suppresses the weight of the desired branch through the accumulation of local survival constraints, while the normalized conditional state preserves the ideal $W_N$ structure.

To benchmark the analytical predictions quantitatively, we also define the corresponding numerical observables in the same symmetric resonant RWA setting.
The numerical many-body state is obtained by direct integration of the interaction-picture Schr\"odinger equation
\begin{equation}
i\hbar \frac{d}{dt}\,|\Psi^{(\mathrm{num})}(t)\rangle
=
\hat H_{I,\mathrm{RWA}}(t)\,
|\Psi^{(\mathrm{num})}(t)\rangle,
\label{eq:TDSE_num_def}
\end{equation}
with the initial condition given by the symmetric $W_N$ atomic resource and the product electron input state introduced above.
The numerical heralding probability is then defined by projection onto the all-ground TLS outcome,
\begin{equation}
P_{G_N}^{(\mathrm{num})}(g,N)
=
\langle\Psi^{(\mathrm{num})}(T)|
\hat{\Pi}_{G_N}
|\Psi^{(\mathrm{num})}(T)\rangle,
\label{eq:PGN_num_def}
\end{equation}
where $\hat{\Pi}_{G_N}=\hat I_e \otimes |G_N\rangle_{\mathrm{TLS}}\,\langle G_N|$ is the projector onto the TLS state with all atoms in the ground state.
Based on Eq.~(\ref{eq:PGN_num_def}), the numerical optimal pulse area is defined as
\begin{equation}
g_{\mathrm{opt}}^{(\mathrm{num})}(N)
=
\operatorname*{arg\,max}_{g}\,
P_{G_N}^{(\mathrm{num})}(g,N),
\label{eq:gopt_num_def}
\end{equation}
and the corresponding numerical maximal heralding probability is
\begin{equation}
P_{G_N}^{\max,(\mathrm{num})}(N)
=
\max_g P_{G_N}^{(\mathrm{num})}(g,N).
\label{eq:PGNmax_num_def}
\end{equation}

\begin{figure}[htpb]
\centering
\scalebox{0.27}{\includegraphics{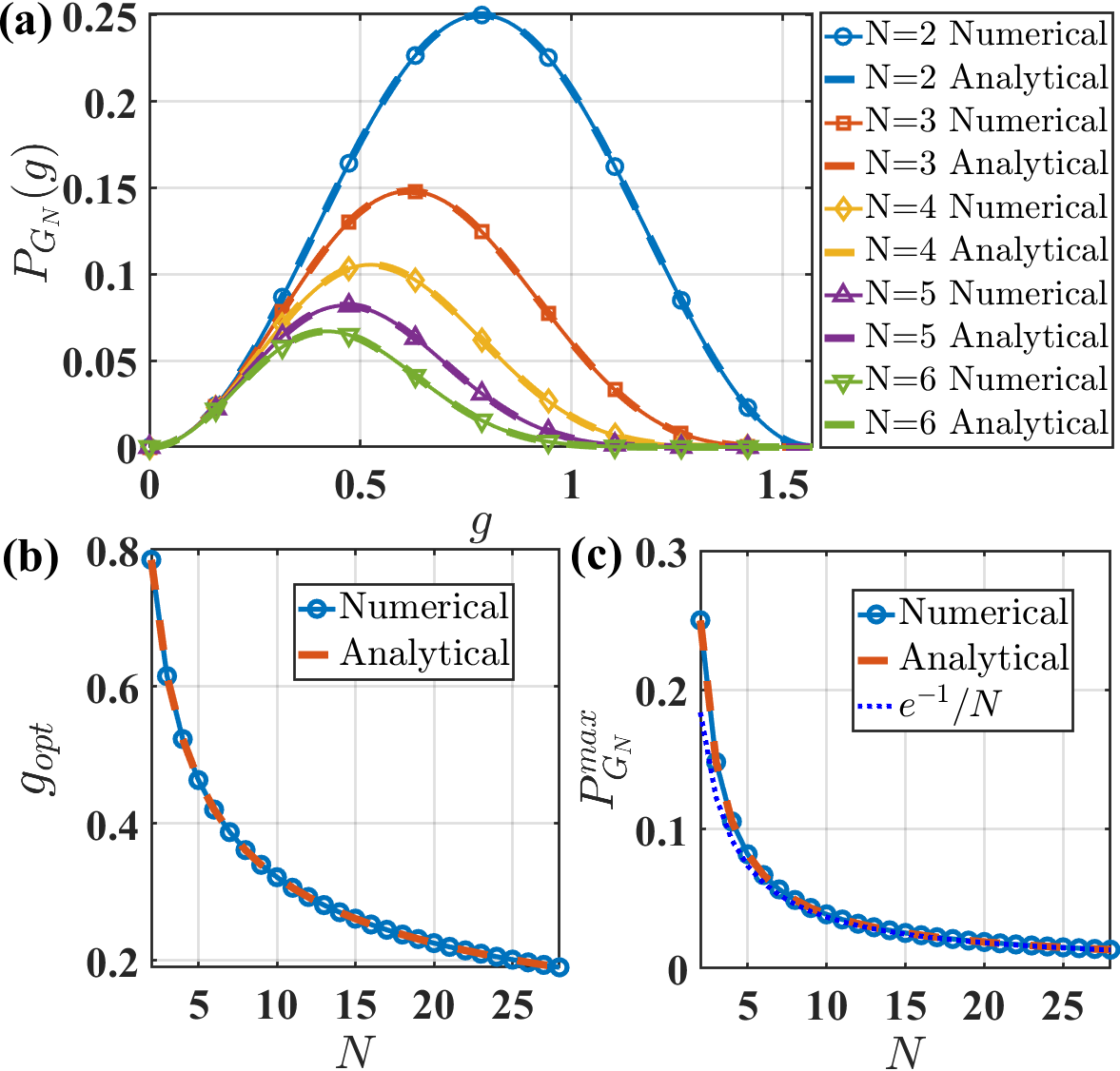}}
\caption{
Validation of the heralded multipartite protocol with the initial atomic resource $c_j=1/\sqrt{N}$.
(a) The heralding probability $P_{G_N}(g)$, where symbols denote numerical results and solid lines denote the analytical expression in Eq.~(\ref{eq:PGN_resonant}).
(b) The optimal pulse area $g_{\mathrm{opt}}$ as a function of $N$, comparing the numerical maxima with the analytical prediction in Eq.~(\ref{eq:g_opt}).
(c) The maximal heralding probability $P_{G_N}^{\max}$ versus $N$, together with the exact analytical result in Eq.~(\ref{eq:PGN_max}) and the asymptotic behavior in Eq.~(\ref{eq:PGN_asymp}).
}
\label{fig:generalN_resonant}
\end{figure}

The analytical results are validated numerically in the symmetric resonant RWA regime.
Figure~\ref{fig:generalN_resonant}(a) compares the numerical heralding probability with the analytical expression in Eq.~(\ref{eq:PGN_resonant}) for several values of $N$.
The numerical data reproduce the full resonant profile, with the peak shifting toward smaller $g$ and decreasing in height as $N$ increases.
The peak shift reflects the growing influence of spectator-arm survival constraints.
Larger $N$ favors operating points that suppress unwanted excitations in the $N-1$ initially unexcited arms, thereby moving the optimum toward weaker effective transfer.
Figure~\ref{fig:generalN_resonant}(b) shows that the numerically extracted optimal pulse area $g_{\mathrm{opt}}$ decreases monotonically with $N$ and agrees quantitatively with Eq.~(\ref{eq:g_opt}).
Figure~\ref{fig:generalN_resonant}(c) shows that the numerical maximum $P_{G_N}^{\max}$ follows the exact result in Eq.~(\ref{eq:PGN_max}) and approaches the large-$N$ asymptotic expression in Eq.~(\ref{eq:PGN_asymp}).

\subsection{Entanglement structure of the heralded state}

We define the upper-sideband target manifold as $\mathcal H_N^{(+)}=\mathrm{span}\{|E_1^{(+)}\rangle,|E_2^{(+)}\rangle,\dots,|E_N^{(+)}\rangle\}$, and the corresponding projector as $\hat P_N^{(+)}=\sum_{j=1}^{N}|E_j^{(+)}\rangle\langle E_j^{(+)}|$.
The subspace $\mathcal H_N^{(+)}$ constitutes the single-excitation electron sector in which one electron occupies the upper sideband and the remaining electrons stay in the reference sideband.
Under the symmetric protocol, $\mathcal H_N^{(+)}$ provides the natural target sector because projection onto the all-ground TLS outcome eliminates every branch with residual TLS excitation and every branch with lower-sideband electron population.

The unconditional electron reduced state is defined as $\hat\rho_e(t)=\mathrm{Tr}_{\mathrm{TLS}}[\,|\Psi(t)\rangle\langle\Psi(t)|\,]$.
The total population weight in the target manifold is given by $w_N^{(+)}(t)=\mathrm{Tr}\bigl[\hat P_N^{(+)}\hat\rho_e(t)\bigr]$.
Equation~(\ref{eq:psi_t_full}) together with the branch structure derived above shows that,   the upper-sideband single-excitation manifold receives contributions only from the TLS outcome $|G_N\rangle_{\mathrm{TLS}}$.
One therefore obtains $w_N^{(+)}(t)=P_{G_N}(t)$, which identifies the target-manifold population directly with the heralding probability.
Population transfer into the desired multipartite electron sector thus occurs only through projection onto the all-ground TLS branch.

The corresponding unconditional fidelity with respect to the ideal multipartite electron target is
\begin{equation}
F_N^{\mathrm{(unc)}}(t)
=
{}_e\langle W_N^{(+)}|
\hat\rho_e(t)
|W_N^{(+)}\rangle_e.
\label{eq:Func_general}
\end{equation}
Because the target manifold is populated only through the heralded branch, one obtains
\begin{equation}
F_N^{\mathrm{(unc)}}(t)=P_{G_N}(t).
\label{eq:Func_equals_P_general}
\end{equation}
All nonheralded branches are orthogonal to the target upper-sideband \(W_N\) manifold after tracing over the TLS sector.
The unconditional figure of merit is therefore controlled entirely by the weight of the desired branch inside the full wavefunction.
The difference between Eqs.~(\ref{eq:Fcond_one_general}) and (\ref{eq:Func_equals_P_general}) is physically significant, i.e., free electron entanglement preparation in the heralded sector is exact, whereas access to that sector remains probabilistic.

Although the heralded target state is genuinely multipartite, pairwise reduced states still offer useful structural insight.
For the ideal heralded state $|W_N^{(+)}\rangle_e$, consider an arbitrary electron pair labeled by $a$ and $b$.
In the effective local basis $\{|0\rangle,|+\rangle\}$ associated with the chosen pair, tracing out the remaining $N-2$ electrons yields the reduced two-electron density matrix
\begin{equation}
\hat\rho_{ab}^{(N)}
=
\frac{N-2}{N}|00\rangle\langle 00|
+\frac{2}{N}|\Psi^+\rangle\langle\Psi^+|,
\label{eq:rhoab_bellmix}
\end{equation}
where $|\Psi^+\rangle=\frac{|+0\rangle+|0+\rangle}{\sqrt 2}$.
The reduced two-electron state consists of a vacuum sector and a symmetric Bell sector.
The vacuum contribution corresponds to configurations in which the transferred excitation resides on electrons other than $a$ or $b$.
The Bell contribution corresponds to configurations in which the transferred excitation resides on electron $a$ or on electron $b$, with coherence preserved between both alternatives.
The reduced two-electron state therefore separates naturally into an ``excitation elsewhere'' sector and an ``excitation coherently shared within the chosen pair'' sector.

For the reduced state $\hat\rho_{ab}^{(N)}$, partial transposition with respect to the second electron gives
\begin{align}
\begin{split}
\hat\rho_{ab}^{(N),\Gamma_b}
=
&\frac{N-2}{N}|00\rangle\langle 00|
+\frac{1}{N}\Bigl(
|+0\rangle\langle +0|
+|0+\rangle\langle 0+|\\
&+|++\rangle\langle 00|
+|00\rangle\langle ++|\Bigr).
\end{split}
\label{eq:rho_PT}
\end{align}
In the ordered basis $\{|00\rangle,|0+\rangle,|+0\rangle,|++\rangle\}$, the matrix representation is
\begin{equation}
\hat\rho_{ab}^{(N),\Gamma_b}
=
\begin{pmatrix}
\frac{N-2}{N} & 0 & 0 & \frac{1}{N}\\
0 & \frac{1}{N} & 0 & 0\\
0 & 0 & \frac{1}{N} & 0\\
\frac{1}{N} & 0 & 0 & 0
\end{pmatrix}.
\label{eq:rho_PT_matrix}
\end{equation}
The corresponding eigenvalues are
\begin{equation}
\lambda_{\pm}=\frac{1}{2N}\left[(N-2)\pm\sqrt{(N-2)^2+4}\right], \
\lambda_3=\lambda_4=\frac{1}{N}.
\label{eq:eigs_PT}
\end{equation}
The negative eigenvalue $\lambda_-$ gives the pairwise negativity
\begin{align}
\mathcal N_{ab}^{(N)}
 =
-\lambda_-
 =
\frac{\sqrt{(N-2)^2+4}-(N-2)}{2N}.
\label{eq:negativity_N}
\end{align}
For large $N$, Eq.~(\ref{eq:negativity_N}) reduces to $\mathcal N_{ab}^{(N)}\sim \frac{1}{N^2}$, which indicates that pairwise entanglement decreases algebraically with increasing electron number, even though the global heralded state remains genuinely multipartite entangled.
Such algebraic decay is a characteristic property of the $W$ class.
A single excitation distributed across $N$ parties supports a finite global correlation structure, whereas any fixed pair retains only a progressively smaller portion of that structure as $N$ increases.

Certification of genuine multipartite entanglement requires a witness beyond pairwise measures.
Introduce the witness operator
\begin{equation}
\hat W_N=\frac{N-1}{N}\hat I-|W_N^{(+)}(\boldsymbol{\phi})\rangle_e\,{}_e\langle W_N^{(+)}(\boldsymbol{\phi})|.
\label{eq:witness_N}
\end{equation}
For every biseparable state $\hat\rho_{\mathrm{bisep}}$, the inequality $\mathrm{Tr}(\hat W_N\hat\rho_{\mathrm{bisep}})\ge 0$ holds, whereas for the ideal heralded target state,
\begin{equation}
\langle\hat W_N\rangle=\mathrm{Tr}\!\left[\hat W_N|W_N^{(+)}(\boldsymbol{\phi})\rangle_e\,{}_e\langle W_N^{(+)}(\boldsymbol{\phi})|\right]
=
-\frac{1}{N}
<0.
\label{eq:witness_ideal}
\end{equation}
It confirms that the ideal heralded state lies outside the biseparable set.
A sufficient condition for genuine multipartite entanglement of a prepared heralded state $\hat\rho_{e|G_N}$ is therefore $\mathrm{Tr}(\hat W_N\hat\rho_{e|G_N})<0$.
The corresponding fidelity threshold is $F_N^{\mathrm{(cond)}}> (N-1)/{N}$.

\subsection{Explicit tripartite example: $N=3$}

\subsubsection{Heralded tripartite free-electron $W$ state}

To make the multipartite structure and the redistribution of correlations explicit, we specialize the general protocol to the tripartite case $N=3$.
The tripartite case provides the minimal genuinely multipartite setting for the heralded transfer protocol.
The initial atomic resource is taken as
\begin{equation}
\begin{split}
|W_3(0)\rangle_{\mathrm{TLS}}
=
\frac{1}{\sqrt 3}
(|egg\rangle
+
|geg\rangle
+
|gge\rangle).
\end{split}
\label{eq:W3_tls}
\end{equation}
The relevant upper-sideband electron basis states are
\begin{align}
|+00\rangle &\equiv |n_0+1,n_0,n_0\rangle, \notag\\
|0+0\rangle &\equiv |n_0,n_0+1,n_0\rangle, \notag\\
|00+\rangle &\equiv |n_0,n_0,n_0+1\rangle.
\label{eq:target_basis_3}
\end{align}
These states specify the three possible arms in which the transferred single excitation can appear in the electron upper-sideband manifold.
Projection of the evolved state onto the all-ground TLS outcome $|G_3\rangle_{\mathrm{TLS}}$ gives
\begin{equation}
\begin{split}
|\Psi_{G_3}^{(e)}(t)\rangle
=
\frac{-i\,s(t)\,[c_-(t)]^2}{\sqrt 3}
(
|+00\rangle
+
|0+0\rangle
+
|00+\rangle
).
\end{split}
\label{eq:psi_ggg_e}
\end{equation}
The factor $-i\,s(t)$ describes the unique transfer event on the arm that initially carries the atomic excitation.
The factor $[c_-(t)]^2$ gives the joint survival amplitude of the remaining two spectator arms.
The corresponding heralding probability is
\begin{equation}
P_{G_3}(t)=|s(t)|^2|c_-(t)|^4,
\label{eq:Pggg_t}
\end{equation}
and the normalized heralded three-electron state is
\begin{equation}
|W_3^{(+)}\rangle_e
=
\frac{1}{\sqrt 3}
\Bigl(
|+00\rangle+|0+0\rangle+|00+\rangle
\Bigr).
\label{eq:W3plus_e}
\end{equation}
The transferred excitation is coherently delocalized over the three electrons.

At the end of a detuned square pulse, the heralding probability follows from Eq.~(\ref{eq:PGN_detuned}) as
\begin{equation}
P_{G_3}(T)
=
\frac{g^2}{\tilde g^2}\sin^2\tilde g
\left(
\cos^2\tilde g+\frac{\delta^2}{\tilde g^2}\sin^2\tilde g
\right)^2.
\label{eq:Pggg_detuned}
\end{equation}
The expression retains the transfer-versus-survival structure of the general-$N$ result, with one transfer event and two spectator-arm survival conditions.
One arm must realize the actual excitation-transfer event, while the remaining two arms must remain in their local ground-sideband sectors.
At resonance, the expression simplifies to
\begin{equation}
P_{G_3}(T)=\sin^2 g\,\cos^4 g.
\label{eq:Pggg_resonant}
\end{equation}
Optimization of Eq.~(\ref{eq:Pggg_resonant}) gives $\cos^2 g_{\mathrm{opt}}  =  {2}/{3}$ and $\sin^2 g_{\mathrm{opt}}  =  {1}/{3}$.
Hence $P_{G_3}^{\max}=4/27\simeq 0.148$.
The tripartite case   displays the generic multipartite mechanism in its simplest nontrivial form, i.e., one successful transfer event accompanied by simultaneous inactivity of the remaining arms.

For $N=3$, the multipartite witness in Eq. (\ref{eq:witness_N}) takes the explicit form
\begin{equation}
\hat W_3=\frac{2}{3}\hat I-|W_3^{(+)}\rangle_e\,{}_e\langle W_3^{(+)}|.
\label{eq:witness_3}
\end{equation}
For the ideal heralded state, the witness expectation value in Eq.(\ref{eq:witness_ideal}) is
\begin{equation}
\langle\hat W_3\rangle=-\frac{1}{3}<0,
\label{eq:wit3}
\end{equation}
which certifies genuine tripartite entanglement and confirms that the heralded state lies outside the biseparable set.
Tracing out one electron, for example electron $3$, gives the reduced state of electrons $1$ and $2$:
\begin{equation}
\hat\rho_{12}^{(3)}
=
\frac{1}{3}|00\rangle\langle 00|
+\frac{2}{3}|\Psi^+\rangle\langle\Psi^+|.
\label{eq:rho12_3}
\end{equation}
The same reduced-state structure holds for $\hat\rho_{13}^{(3)}$ and $\hat\rho_{23}^{(3)}$ by symmetry.
The corresponding negativity is
\begin{equation}
\mathcal N_{12}^{(3)}=\mathcal N_{13}^{(3)}=\mathcal N_{23}^{(3)}=\frac{\sqrt 5-1}{6}\simeq 0.206.
\label{eq:N_pairs_3}
\end{equation}
It shows that every two-electron reduction retains nonzero bipartite entanglement.
The result is characteristic of the $W$ class, in which delocalization of a single excitation over all parties leaves entanglement visible in every pairwise reduction.

\begin{figure}[htpb]
\centering
\scalebox{0.27}{\includegraphics{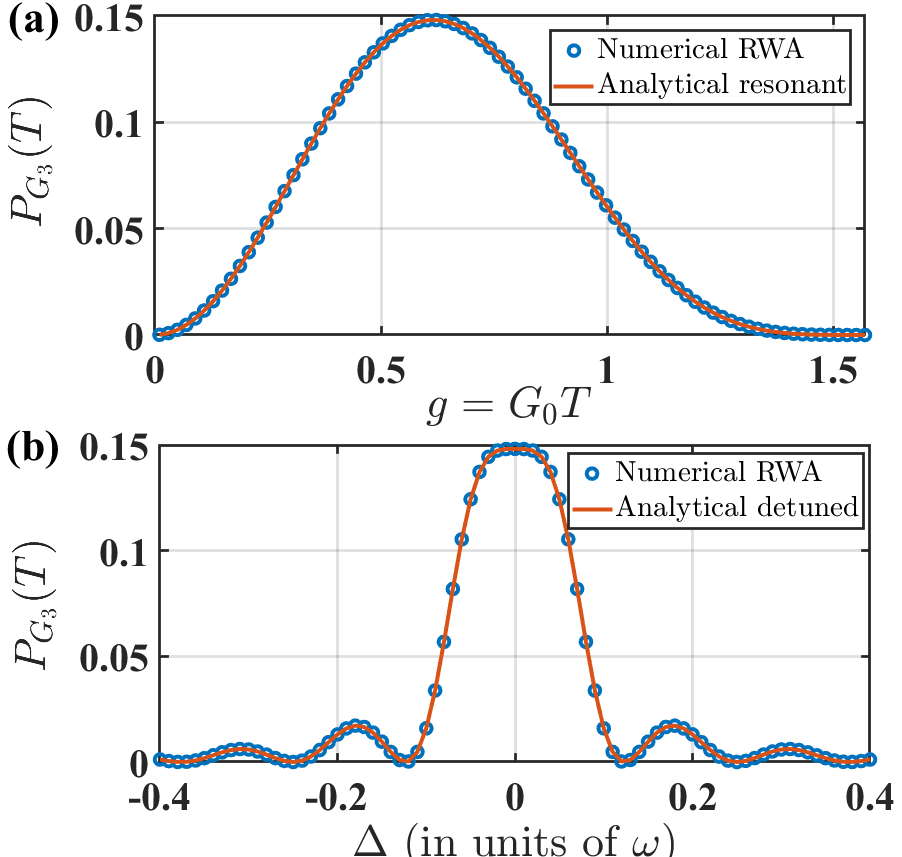}}
\caption{
Tripartite numerical validation for $N=3$.
(a) The numerical resonant heralding probability with the analytical result $P_{G_3}(T)$ in Eq.~(\ref{eq:Pggg_resonant}) as a function of the dimensionless pulse area $g=G_0T$.
The curve rises from zero, reaches a single optimum $P_{G_3}^{\max}$ determined by the balance between one-arm transfer and two-arm survival, and then decreases as stronger driving enhances unwanted dynamics on the spectator arms.
(b) The numerical and analytical detuned heralding probabilities at fixed pulse area as a function of the normalized detuning.
The probability is maximal near resonance and decreases symmetrically as the detuning magnitude grows, reflecting the reduced efficiency of coherent excitation exchange under phase mismatch.
}
\label{fig:N3_validation}
\end{figure}

The tripartite analytical results admit direct numerical validation within the symmetric RWA dynamics.
Figure~\ref{fig:N3_validation}(a) shows the resonant heralding probability $P_{G_3}(T)$ as a function of the dimensionless pulse area $g=G_0T$.
The numerical data coincide with the analytical expression in Eq.~(\ref{eq:Pggg_resonant}), confirming the tripartite specialization of the general theory.
The profile of $P_{G_3}(T)$ has a transparent physical interpretation.
For small $g$, the transfer amplitude is weak and the heralded branch is strongly suppressed.
As $g$ increases, the transfer channel becomes stronger and the heralding probability rises.
Beyond the optimum $P_{G_3}^{\max}=  {4}/{27}$, stronger local driving then enhances unwanted dynamics on the two spectator arms, so the probability decreases.
The nonmonotonic shape therefore reflects the same transfer-versus-survival competition identified in the general-$N$ analysis, now specialized to one transfer arm and two spectator arms.

Figure~\ref{fig:N3_validation}(b) shows the detuned tripartite heralding probability as a function of the normalized detuning.
The numerical data follow the analytical detuned expression in Eq.~(\ref{eq:Pggg_detuned}), confirming that finite detuning primarily changes the success probability while leaving the heralded-state structure intact in the symmetric configuration.
A finite $\Delta$ weakens resonant excitation exchange and therefore suppresses the branch in which the initial TLS excitation is converted into an electron upper-sideband excitation.
At the same time, detuning modifies the survival amplitudes of the spectator arms through the generalized Rabi factors $c_-(T)$ and $s(T)$.
The resulting suppression is symmetric with respect to the sign of $\Delta$, consistent with the fact that only the magnitude of the mismatch enters the present symmetric protocol.
The results show that the tripartite benchmark is fully captured by the analytical RWA theory at the level of both pulse-area dependence and detuning dependence.

\subsubsection{Time-resolved entanglement redistribution}

For the symmetric tripartite protocol, a time-resolved analysis is useful for distinguishing among three notions of electron entanglement, namely, the intrinsic entanglement of the full electron subsystem, the ideal entanglement structure of the heralded branch, and the practically accessible entanglement yield after postselection.
The numerical time evolution is obtained from the Schr\"odinger equation (\ref{eq:TDSE_num_def}).
For the evolved pure state $|\Psi^{(\mathrm{num})}(t)\rangle$, the atomic reduced state is
\begin{equation}
\hat\rho^{(\mathrm{num})}_A(t)=\mathrm{Tr}_e\bigl[|\Psi^{(\mathrm{num})}(t)\rangle\langle\Psi^{(\mathrm{num})}(t)|\bigr],
\label{eq:rhoA_t}
\end{equation}
For any chosen atomic pair $(a,b)\in\{(1,2),(1,3),(2,3)\}$, the corresponding two-TLS reduced state is
\begin{equation}
\hat\rho^{\,A}_{ab}(t)
=
\mathrm{Tr}_{c}\!\bigl[\hat\rho^{(\mathrm{num})}_A(t)\bigr],\
\{a,b,c\}=\{1,2,3\},
\label{eq:rhoA_ab_t}
\end{equation}
and the pairwise atomic entanglement is quantified by the negativity
\begin{equation}
\mathcal N^{A}_{ab}(t)
=
\frac{
\left\|
\bigl(\hat\rho^{\,A}_{ab}(t)\bigr)^{\Gamma_b}
\right\|_1-1
}{2},
\label{eq:NA_ab_t}
\end{equation}
where $\Gamma_b$ denotes partial transposition with respect to subsystem $b$, and $\|\cdot\|_1$ is the trace norm.
The atomic average pairwise negativity is then evaluated numerically as
\begin{equation}
\bar{\mathcal N}^{A}_{\mathrm{pair}}(t)
=
\frac{
\mathcal N^{A}_{12}(t)+\mathcal N^{A}_{13}(t)+\mathcal N^{A}_{23}(t)
}{3}.
\label{eq:NA_pair_t}
\end{equation}

The unconditional electron reduced state is
\begin{equation}
\hat\rho^{(\mathrm{num})}_e(t)=\mathrm{Tr}_A\bigl[|\Psi^{(\mathrm{num})}(t)\rangle\langle\Psi^{(\mathrm{num})}(t)|\bigr],
\label{eq:rhoe_t}
\end{equation}
where the trace is taken over the three TLS degrees of freedom.
For any chosen electron pair $(a,b)\in\{(1,2),(1,3),(2,3)\}$, the corresponding two-electron reduced state is
\begin{equation}
\hat\rho^{\,e}_{ab}(t)
=
\mathrm{Tr}_{c}\!\bigl[\hat\rho^{(\mathrm{num})}_e(t)\bigr],\
\{a,b,c\}=\{1,2,3\},
\label{eq:rhoe_ab_t}
\end{equation}
and the pairwise electron entanglement is quantified by the negativity
\begin{equation}
\mathcal N^{e}_{ab}(t)
=
\frac{
\left\|
\bigl(\hat\rho^{\,e}_{ab}(t)\bigr)^{\Gamma_b}
\right\|_1-1
}{2},
\label{eq:Ne_ab_t}
\end{equation}
where $\Gamma_b$ denotes partial transposition with respect to electron $b$, and $\|\cdot\|_1$ is the trace norm.
The time-resolved average pairwise entanglement of the unconditional electron subsystem is therefore
\begin{equation}
\bar{\mathcal N}^{\,e}_{\mathrm{pair}}(t)
=
\frac{
\mathcal N^{e}_{12}(t)+\mathcal N^{e}_{13}(t)+\mathcal N^{e}_{23}(t)
}{3}.
\label{eq:Ne_pair_t}
\end{equation}
In the fully symmetric tripartite protocol, the reduced electron state remains permutation symmetric under exchange of the three electron arms.
Consequently, $\mathcal N^{e}_{12}(t)=\mathcal N^{e}_{13}(t)=\mathcal N^{e}_{23}(t)$, so that Eq.~(\ref{eq:Ne_pair_t}) reduces to $\bar{\mathcal N}^{\,e}_{\mathrm{pair}}(t)=\mathcal N^{e}_{12}(t)$.
The quantity $\bar{\mathcal N}^{\,e}_{\mathrm{pair}}(t)$ is determined by the full electron reduced state and therefore contains contributions from both the heralded branch and all nonheralded branches.
The numerical heralding probability associated with the TLS outcome $|G_3\rangle_{TLS}$ is
\begin{equation}
P_{G_3}^{(\mathrm{num})}(t)
=
\langle\Psi^{(\mathrm{num})}(t)|
\bigl(\hat I_e\otimes |G_3\rangle_{TLS}\langle G_3|\bigr)
|\Psi^{(\mathrm{num})}(t)\rangle,
\label{eq:Pggg_num_t}
\end{equation}
and the corresponding postselected electron state is
\begin{equation}
\hat\rho^{(\mathrm{num})}_{e|G_3}(t)
=
\frac{
{}_{\mathrm{TLS}}\langle G_3|\Psi^{(\mathrm{num})}(t)\rangle
\langle\Psi^{(\mathrm{num})}(t)|G_3\rangle_{\mathrm{TLS}}
}{
P_{G_3}^{(\mathrm{num})}(t)
}.
\label{eq:rhoe_cond_num_t}
\end{equation}
For any chosen electron pair $(a,b)\in\{(1,2),(1,3),(2,3)\}$, the corresponding conditional two-electron reduced state is
\begin{equation}
\hat\rho^{\,e,\mathrm{cond}}_{ab}(t)
=
\mathrm{Tr}_{c}\!\bigl[\hat\rho^{(\mathrm{num})}_{e|ggg}(t)\bigr],\
\{a,b,c\}=\{1,2,3\},
\label{eq:rhoe_cond_ab_t}
\end{equation}
and the conditional pairwise electron entanglement is quantified by
\begin{equation}
\mathcal N^{e,\mathrm{cond}}_{ab}(t)
=
\frac{
\left\|
\bigl(\hat\rho^{\,e,\mathrm{cond}}_{ab}(t)\bigr)^{\Gamma_b}
\right\|_1-1
}{2}.
\label{eq:Ne_cond_ab_t}
\end{equation}
The numerical conditional electron average pairwise negativity is therefore
\begin{equation}
\bar{\mathcal N}^{\,e,\mathrm{cond}}_{\mathrm{pair}}(t)
=
\frac{
\mathcal N^{e,\mathrm{cond}}_{12}(t)+
\mathcal N^{e,\mathrm{cond}}_{13}(t)+
\mathcal N^{e,\mathrm{cond}}_{23}(t)
}{3},
\label{eq:Ne_cond_pair_num_t}
\end{equation}
and the corresponding numerical success-weighted conditional electron entanglement yield is defined as
\begin{equation}
Y_{\mathrm{pair}}^{(\mathrm{num})}(t)
=
P_{G_3}^{(\mathrm{num})}(t)\,
\bar{\mathcal N}^{\,e,\mathrm{cond}}_{\mathrm{pair}}(t).
\label{eq:Y_pair_num_t}
\end{equation}

By contrast, the conditional electron pairwise entanglement admits a closed-form analytical benchmark.
Postselection of the TLS outcome $|G_3\rangle_{TLS}$ projects the electron subsystem onto the ideal three-electron $W_3$-type state within the symmetric RWA treatment.
The conditional electron average pairwise negativity is therefore time independent and equals
\begin{equation}
\mathcal N^{e,\mathrm{cond}}_{\mathrm{pair}}(t)
=
\frac{\sqrt5-1}{6}\simeq 0.206.
\label{eq:Ne_cond_pair_const}
\end{equation}
which provides the analytical benchmark.
Strictly speaking, at times for which $P_{G_3}(t)=0$ the conditional state is not operationally accessible, but the constant value in Eq.~(\ref{eq:Ne_cond_pair_const}) remains the correct entanglement content of the ideal heralded branch whenever that branch is successfully selected.

Another useful quantity is the success-weighted conditional electron pairwise entanglement, $Y_{\mathrm{pair}}(t)=P_{G_3}(t)\,\mathcal N^{e,\mathrm{cond}}_{\mathrm{pair}}(t)$,
which characterizes the practically accessible electron entanglement yield after postselection.
In the symmetric detuned RWA case, $P_{G_3}(t)=|s(t)|^2|c_-(t)|^4$, so that
\begin{equation}
Y_{\mathrm{pair}}(t)
=
\frac{\sqrt5-1}{6}\,|s(t)|^2|c_-(t)|^4.
\label{eq:Y_pair_detuned}
\end{equation}
At exact resonance, Eq.~(\ref{eq:Y_pair_detuned}) reduces to
\begin{equation}
Y_{\mathrm{pair}}(t)
=
\frac{\sqrt5-1}{6}\,\sin^2 g(t)\cos^4 g(t).
\label{eq:Y_pair_resonant}
\end{equation}

\begin{figure}[htpb]
\centering
\scalebox{0.33}{\includegraphics{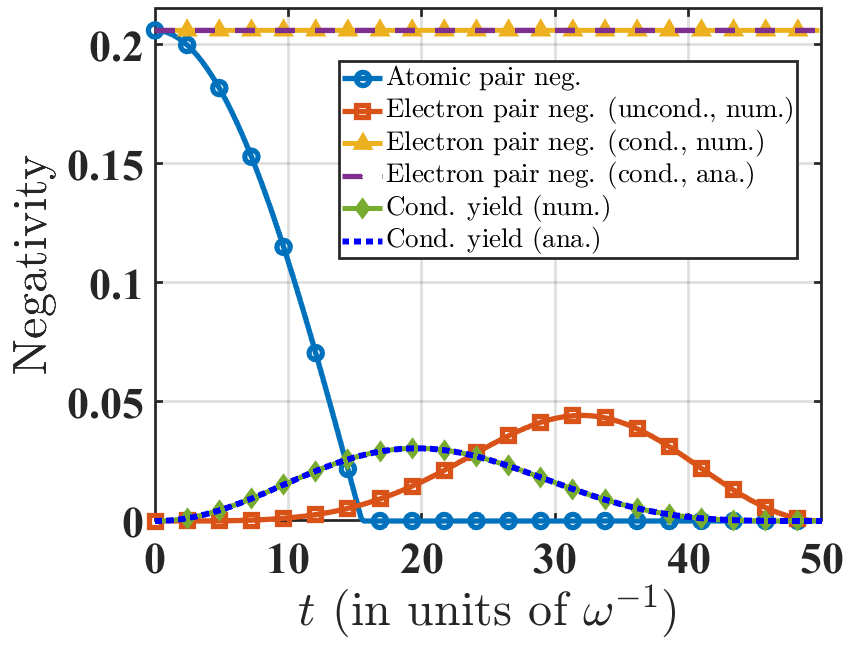}}
\caption{
Time-resolved entanglement redistribution in the symmetric resonant tripartite protocol for $N=3$, starting from the symmetric atomic resource $|W_3\rangle_{\mathrm{TLS}}$.
The solid blue curve denotes the numerical atomic average pairwise negativity defined in Eq.~(\ref{eq:NA_pair_t}).
The solid orange curve denotes the numerical unconditional electron average pairwise negativity defined in Eq.~(\ref{eq:Ne_pair_t}).
The solid yellow curve with markers denotes the numerical conditional electron average pairwise negativity defined in Eq.~(\ref{eq:Ne_cond_pair_num_t}), and the dashed purple curve denotes the analytical result in Eq.~(\ref{eq:Ne_cond_pair_const}).
The solid green curve with markers denotes the numerical success-weighted conditional electron entanglement yield defined in Eq.~(\ref{eq:Y_pair_num_t}), and the dotted blue curve denotes the analytical result in Eq.~(\ref{eq:Y_pair_resonant}).
Parameters: $\Delta=0$ and $g=\pi/2$.
}
\label{fig:N3_AFt}
\end{figure}

Figure~\ref{fig:N3_AFt} presents the numerical atomic average pairwise entanglement $\bar{\mathcal N}^{A}_{\mathrm{pair}}(t)$ in Eq.~(\ref{eq:NA_pair_t}) together with the numerical unconditional electron average pairwise entanglement $\bar{\mathcal N}^{\,e}_{\mathrm{pair}}(t)$ in Eq.~(\ref{eq:Ne_pair_t}), providing a direct view of entanglement redistribution between the atomic and electron subsystems.
Figure~\ref{fig:N3_AFt} also compares the numerical conditional electron average pairwise entanglement $\bar{\mathcal N}^{\,e,\mathrm{cond}}_{\mathrm{pair}}(t)$ in Eq.~(\ref{eq:Ne_cond_pair_num_t}) with the analytical result in Eq.~(\ref{eq:Ne_cond_pair_const}), and the numerical success-weighted conditional electron entanglement yield $Y_{\mathrm{pair}}^{(\mathrm{num})}(t)$ in Eq.~(\ref{eq:Y_pair_num_t}) with the analytical result in Eq.~(\ref{eq:Y_pair_resonant}).
The average atomic pairwise negativity $\bar{\mathcal N}^{A}_{\mathrm{pair}}(t)$ decreases monotonically during the evolution and eventually vanishes, indicating progressive depletion of the initial atomic pairwise entanglement.
The average electron pairwise negativity $\bar{\mathcal N}^{\,e}_{\mathrm{pair}}(t)$ starts from zero, increases as correlations build up in the electron sector, reaches a maximum, and decreases at later times.
This behavior arises because the reduced state in Eq.~(\ref{eq:rhoe_t}) contains contributions from all dynamical branches, not only from the ideal heralded branch.
Unconditional electron entanglement is therefore generated gradually and shaped by the competition between coherent transfer and branch mixing.

The numerical conditional electron average pairwise negativity $\bar{\mathcal N}^{\,e,\mathrm{cond}}_{\mathrm{pair}}(t)$ in Eq.~(\ref{eq:Ne_cond_pair_num_t}) remains constant throughout the evolution and coincides with the analytical value $\mathcal N^{e,\mathrm{cond}}_{\mathrm{pair}}(t)$ in Eq.~(\ref{eq:Ne_cond_pair_const}), which shows that the conditioned electron state preserves the ideal tripartite $W_3$ structure once the heralding condition is imposed.
Agreement between the numerical and analytical conditional curves confirms that, within the symmetric RWA treatment, the branch selected by the TLS outcome retains the exact ideal multipartite form at all times.
The practically accessible postselected entanglement is characterized by $Y_{\mathrm{pair}}^{(\mathrm{num})}(t)$ in Eq.~(\ref{eq:Y_pair_num_t}), while the analytical benchmark is given by $Y_{\mathrm{pair}}(t)$ in Eq.~(\ref{eq:Y_pair_resonant}).
The numerical yield rises from zero, reaches a maximum near the optimal heralding point, and then decreases as the heralding probability is suppressed at later times.
The peak position of $Y_{\mathrm{pair}}^{(\mathrm{num})}(t)$ precedes the maximum of $\bar{\mathcal N}^{\,e}_{\mathrm{pair}}(t)$, which indicates that the time at which the ideal heralded branch is most accessible does not coincide with the time at which the full electron subsystem carries the largest intrinsic pairwise entanglement.
The quantity $\bar{\mathcal N}^{\,e}_{\mathrm{pair}}(t)$ therefore characterizes intrinsic pairwise entanglement in the full electron subsystem, whereas $\bar{\mathcal N}^{\,e,\mathrm{cond}}_{\mathrm{pair}}(t)$ characterizes the ideal entanglement carried by the heralded branch and $Y_{\mathrm{pair}}^{(\mathrm{num})}(t)$ quantifies the accessible postselected entanglement yield.

\subsubsection{Entanglement transfer of weighted single-excitation states}

To quantify how the entanglement content of a general single-excitation atomic resource is transferred to the free-electron subsystem, we consider the weighted tripartite state
\begin{equation}
|\Phi_A(0)\rangle
=
\alpha |egg\rangle
+
\beta |geg\rangle
+
\gamma |gge\rangle,
\label{eq:weighted_tripartite_input}
\end{equation}
where $|\alpha|^2+|\beta|^2+|\gamma|^2=1$.
The amplitudes are parametrized as
\begin{equation}
\alpha=\cos\theta,
\
\beta=\sin\theta\cos\phi,
\
\gamma=\sin\theta\sin\phi,
\label{eq:theta_phi_param}
\end{equation}
with $0\le \theta \le \frac{\pi}{2}$ and $0\le \phi \le \frac{\pi}{2}$.

Under the symmetric resonant RWA protocol, the heralded three-electron branch preserves the weighted single-excitation structure exactly, as given in Eq. (\ref{eq:Fcond_one_general}).
Postselection of the TLS outcome $|G_3\rangle_{TLS}$ yields the normalized electron state
\begin{equation}
|\psi_e^{(\mathrm{cond})}(T)\rangle
=
\alpha |+00\rangle
+
\beta |0+0\rangle
+
\gamma |00+\rangle.
\label{eq:weighted_tripartite_output}
\end{equation}
It shows that the protocol realizes an isometric mapping from the initial atomic single-excitation manifold to the heralded electron upper-sideband manifold.
For pairwise correlations, define the initial atomic average pairwise negativity as
\begin{equation}
\bar{\mathcal N}^{A}_{\mathrm{pair}}(0)
=
\frac{
\mathcal N^{A}_{12}(0)
+
\mathcal N^{A}_{13}(0)
+
\mathcal N^{A}_{23}(0)
}{3},
\label{eq:avg_pair_neg_atomic}
\end{equation}
and the heralded electron average pairwise negativity as
\begin{equation}
\bar{\mathcal N}^{e,\mathrm{cond}}_{\mathrm{pair}}(T)
=
\frac{
\mathcal N^{e,\mathrm{cond}}_{12}(T)
+
\mathcal N^{e,\mathrm{cond}}_{13}(T)
+
\mathcal N^{e,\mathrm{cond}}_{23}(T)
}{3}.
\label{eq:avg_pair_neg_electron}
\end{equation}
For the one-versus-rest bipartite structure of the initial three-TLS resource, we use the von Neumann entropy
\begin{equation}
S^{A}_{1|23}(0)
=
-
\mathrm{Tr}
\bigl[
\rho^{A}_{1}(0)\log_2\rho^{A}_{1}(0)
\bigr],
\label{eq:S_atomic_123}
\end{equation}
where $\rho^{A}_{1}(0)$ is the reduced density matrix of TLS $1$ obtained by tracing over TLS $2$ and TLS $3$.
For the one-versus-rest bipartite structure of the heralded three-electron state, we use the von Neumann entropy
\begin{equation}
S^{e,\mathrm{cond}}_{1|23}(T)
=
-
\mathrm{Tr}
\bigl[
\rho^{e,\mathrm{cond}}_{1}(T)\log_2\rho^{e,\mathrm{cond}}_{1}(T)
\bigr].
\label{eq:S_electron_123}
\end{equation}
Here $\rho^{e,\mathrm{cond}}_{1}(T)$ denotes the reduced density matrix of electron $1$ in the conditional output state after tracing over electrons $2$ and $3$.
Exact preservation of the amplitudes $(\alpha,\beta,\gamma)$ in Eq.~(\ref{eq:weighted_tripartite_output}) leads to the analytical relations
\begin{equation}
\bar{\mathcal N}^{e,\mathrm{cond}}_{\mathrm{pair}}(T)
=
\bar{\mathcal N}^{A}_{\mathrm{pair}}(0),
\label{eq:pairwise_transfer_identity}
\end{equation}
and
\begin{equation}
S^{e,\mathrm{cond}}_{1|23}(T)
=
S^{A}_{1|23}(0).
\label{eq:entropy_transfer_identity}
\end{equation}

The heralding probability, by contrast, is determined only by the dynamical coefficients and is independent of the weight distribution $(\alpha,\beta,\gamma)$.
For the symmetric tripartite protocol, $P_{G_3}(T)=|s(T)|^2|c_-(T)|^4$,
which reduces at exact resonance in Eq. (\ref{eq:Pggg_resonant}).
At the optimal pulse area $g_{\mathrm{opt}}=\arccos\sqrt{\frac{2}{3}}$,
the success probability becomes
\begin{equation}
P_{G_3} (T)=\frac{4}{27}\simeq 0.148,
\label{eq:Pgggmax_weighted}
\end{equation}
independent of the initial weighted $W_3$ resource.
The conditional fidelity with respect to the target heralded electron state is defined as
\begin{equation}
F^{(\mathrm{cond})}_{3}
=
\langle
\Phi_e^{(\mathrm{target})}
|
\hat{\rho}_e^{(\mathrm{cond})}(T)
|
\Phi_e^{(\mathrm{target})}
\rangle,
\label{eq:Fcond_weighted}
\end{equation}
where  $|\Phi_e^{(\mathrm{target})}\rangle=\alpha |+00\rangle+\beta |0+0\rangle+\gamma |00+\rangle$ and $\hat{\rho}_e^{(\mathrm{cond})}(T)=|\psi_{G_3}^{(\mathrm{e})}(T)\rangle\langle \psi_{G_3}^{(\mathrm{e})}(T)|$.
Within the ideal symmetric RWA treatment,
\begin{equation}
F^{(\mathrm{cond})}_{3}=1.
\label{eq:Fcond_weighted_one}
\end{equation}

\begin{figure}[htpb]
\centering
\scalebox{0.235}{\includegraphics{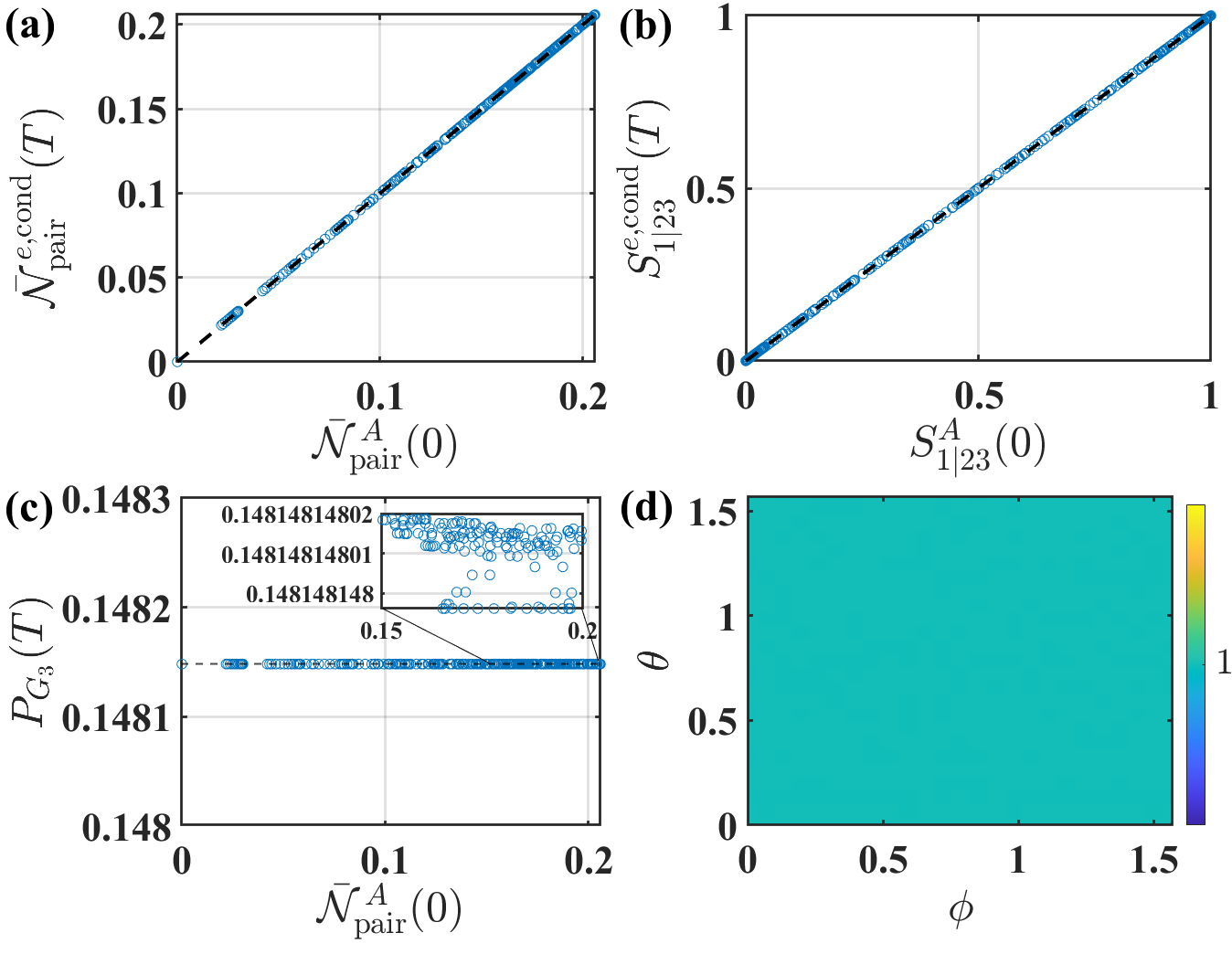}}
\caption{
Transfer of weighted single-excitation entanglement in the symmetric resonant tripartite protocol for $N=3$.
The family of initial weighted states is given by Eq.~(\ref{eq:weighted_tripartite_input}) and parametrized by Eq.~(\ref{eq:theta_phi_param}) over $0\le\theta\le\pi/2$ and $0\le\phi\le\pi/2$.
(a) The initial atomic average pairwise negativity in Eq.~(\ref{eq:avg_pair_neg_atomic}) vs the corresponding heralded electron average pairwise negativity in Eq.~(\ref{eq:avg_pair_neg_electron}), together with the analytical relation in Eq.~(\ref{eq:pairwise_transfer_identity}).
(b) The initial atomic one-versus-rest bipartite entropy in Eq.~(\ref{eq:S_atomic_123}) vs the corresponding heralded electron entropy in Eq.~(\ref{eq:S_electron_123}), together with the analytical relation in Eq.~(\ref{eq:entropy_transfer_identity}).
(c) The heralding probability as a function of the initial atomic average pairwise entanglement at the optimal pulse area $g_{\mathrm{opt}}=\arccos\sqrt{2/3}$, with the analytical value given by Eq.~(\ref{eq:Pgggmax_weighted}).
(d) The conditional fidelity as a function of the scanned weighted initial states at $g_{\mathrm{opt}}=\arccos\sqrt{2/3}$, with the ideal analytical value given by Eq.~(\ref{eq:Fcond_weighted_one}).
}
\label{fig:weighted_resource_scan}
\end{figure}

Figure~\ref{fig:weighted_resource_scan}(a) places the initial atomic average pairwise negativity in Eq.~(\ref{eq:avg_pair_neg_atomic}) against the heralded electron average pairwise negativity in Eq.~(\ref{eq:avg_pair_neg_electron}).
Numerical points remain on the diagonal relation specified by Eq.~(\ref{eq:pairwise_transfer_identity}).
Figure~\ref{fig:weighted_resource_scan}(b) places the initial atomic one-versus-rest entropy in Eq.~(\ref{eq:S_atomic_123}) against the heralded electron one-versus-rest entropy in Eq.~(\ref{eq:S_electron_123}).
Numerical points again follow the diagonal relation given by Eq.~(\ref{eq:entropy_transfer_identity}).
Such behavior indicates that the weighted single-excitation structure is preserved in the heralded branch at the level of both average pairwise negativity and one-versus-rest bipartite entropy.
Preservation of the amplitudes in Eq.~(\ref{eq:weighted_tripartite_output}) provides the underlying reason, since the corresponding bipartite spectra remain unchanged under the symmetric resonant mapping.

Figure~\ref{fig:weighted_resource_scan}(c) shows the optimal heralding probability as a function of the initial atomic average pairwise negativity defined in Eq.~(\ref{eq:avg_pair_neg_atomic}).
In the numerical scan, the weighted $W_3$ resource is parametrized by the two angles $(\theta,\phi)$, whereas the horizontal coordinate in Fig.~\ref{fig:weighted_resource_scan}(c) is the single scalar quantity $\bar{\mathcal N}^{A}_{\mathrm{pair}}(0)$.
Different pairs of $(\theta,\phi)$ can therefore yield the same value of $\bar{\mathcal N}^{A}_{\mathrm{pair}}(0)$, so several distinct input resources may be projected onto the same horizontal coordinate.
The corresponding numerical samples are plotted separately, which explains the occurrence of multiple points at the same value of $\bar{\mathcal N}^{A}_{\mathrm{pair}}(0)$.
Within the symmetric resonant theory, the optimal heralding probability is fixed by Eq.~(\ref{eq:Pgggmax_weighted}) and does not depend on the detailed weight distribution of the initial resource.
The small vertical spread visible in the inset is therefore attributed to numerical resolution and finite-precision effects rather than to a physical multivalued dependence of $P_{G_3}(T)$ on $\bar{\mathcal N}^{A}_{\mathrm{pair}}(0)$.
Figure~\ref{fig:weighted_resource_scan}(d) shows the conditional fidelity defined in Eq.~(\ref{eq:Fcond_weighted}), which remains at the analytical benchmark in Eq.~(\ref{eq:Fcond_weighted_one}) throughout the scanned $(\theta,\phi)$ region.
The results confirm that the symmetric resonant dynamics fixes the success probability through the transfer and survival factors, while the weight distribution of the input resource is faithfully inherited by the heralded weighted free-electron $W_3$ state.

\section{Discussion}\label{DISS}

\subsection{Robustness to detuning and symmetry breaking}

\begin{figure}[htpb]
\centering
\scalebox{0.235}{\includegraphics{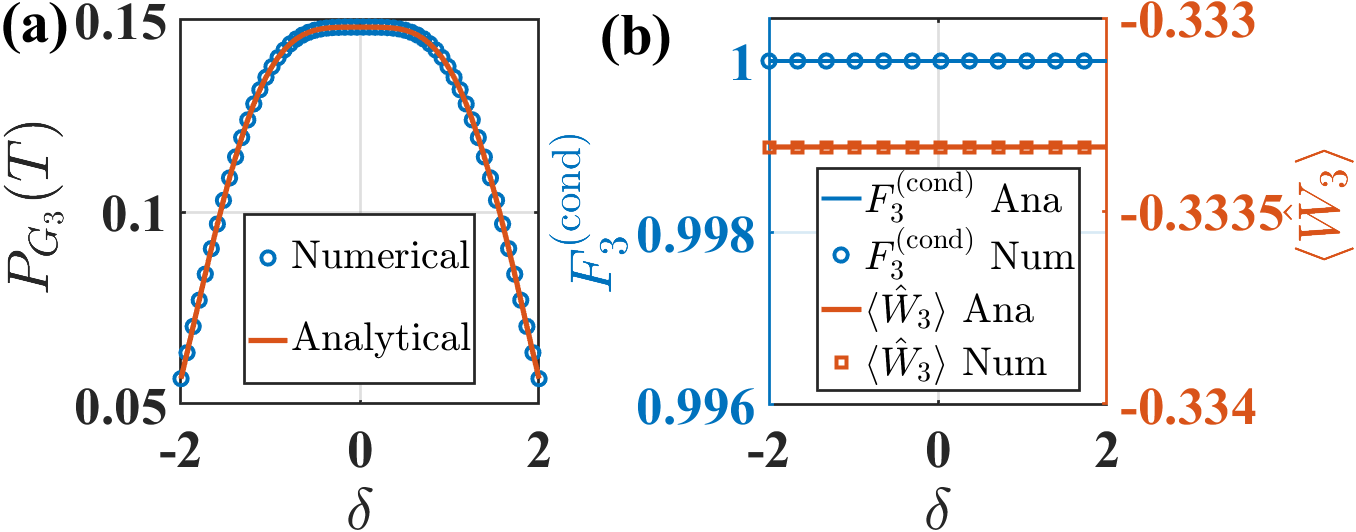}}
\caption{
Common-detuning robustness of the symmetric tripartite protocol for $N=3$.
A common detuning is introduced in an otherwise symmetric configuration, with $\delta=\Delta T/2$.
(a) The heralding probability $P_{G_3}(T)$ as a function of the normalized detuning $\delta$, with the analytical result given by Eq.~(\ref{eq:PGN_detuned});
(b) The conditional fidelity $F_3^{(\mathrm{cond})}(T)$ associated with Eq.~(\ref{eq:Fcond_one_general}) and the witness expectation value associated with the tripartite witness operator in Eq.~(\ref{eq:witness_3}) as functions of $\delta$.
Parameters: $g=\arccos\sqrt{2/3}$ and $\Delta_1=\Delta_2=\Delta_3=\Delta$.
}
\label{fig:dis_common_detuning}
\end{figure}

We now examine how common detuning and weak symmetry breaking affect the multipartite transfer protocol.
Common detuning is introduced within an otherwise symmetric configuration, whereas weak symmetry breaking is modeled by small arm-dependent deviations in the local couplings.
We begin with the symmetric detuned case.
For $N=3$, Eq.~(\ref{eq:PGN_detuned}) predicts a heralding probability that is maximal near resonance and decreases as $|\Delta|$ increases, while Eq.~(\ref{eq:Fcond_one_general}) implies that the conditional fidelity remains unchanged.
Figure~\ref{fig:dis_common_detuning}(a) shows excellent agreement between direct simulations of $P_{G_3}(T)$ and the analytical expression in Eq.~(\ref{eq:PGN_detuned}) over the detuning scan.
In contrast, Fig.~\ref{fig:dis_common_detuning}(b) shows that both the conditional fidelity $F_3^{(\mathrm{cond})}$ and the witness expectation value remain essentially constant over the same range.
The numerical results therefore highlight a clear physical distinction.
Common detuning suppresses the probability of accessing the heralded branch, whereas the symmetry of the interaction preserves the entanglement structure of the postselected electron state once the branch is obtained.

We next turn to weak symmetry breaking between different local arms.
In realistic implementations, such imperfections may arise from small variations in coupling strength, interaction duration, or local detuning.
The postselected state then remains confined to the upper-sideband single-excitation manifold, but the ideal symmetric $W_N$ state is replaced by a weighted superposition.
For the $j$th arm, we define
\begin{align}
s_j(T) &= \frac{g_j}{\tilde g_j}\sin\tilde g_j,  \
c_{j,-}(T) &= \cos\tilde g_j-i\frac{\delta_j}{\tilde g_j}\sin\tilde g_j,
\label{eq:sj_cj_disc}
\end{align}
with $g_j = G_{0,j}T$, $\delta_j = \frac{\Delta_jT}{2}$ and $\tilde g_j = \sqrt{g_j^2+\delta_j^2}$.
After postselecting the TLS outcome $|G_N\rangle_{\mathrm{TLS}}$, the unnormalized heralded electron state becomes
\begin{equation}
|\Psi_{G_N}^{(e)}(T)\rangle
=
\frac{-i}{\sqrt N}
\sum_{j=1}^{N}
e^{i\phi_j}
s_j(T)
\prod_{k\neq j}c_{k,-}(T)
|E_j^{(+)}\rangle.
\label{eq:weighted_unnorm_disc}
\end{equation}
Introducing
\begin{equation}
\alpha_j(T)
=
e^{i\phi_j}
s_j(T)
\prod_{k\neq j}c_{k,-}(T),
\label{eq:alpha_j_disc}
\end{equation}
the normalized heralded state can be written as
\begin{equation}
|\widetilde W_N^{(+)}(T)\rangle_e
=
\frac{\sum_{j=1}^{N}\alpha_j(T)|E_j^{(+)}\rangle}
{\sqrt{\sum_{j=1}^{N}|\alpha_j(T)|^2}}.
\label{eq:weighted_state_disc}
\end{equation}
The equation  shows that weak symmetry breaking does not destroy the single-excitation multipartite character of the transfer.
The output still lies in the target upper-sideband manifold.
The nonideal effect appears instead through the relative amplitudes $\alpha_j(T)$, which distort the equal-weight interference structure of the ideal $W_N$ state.

A convenient measure of this deformation is the conditional fidelity with respect to the ideal symmetric target,
\begin{equation}
F_N^{\mathrm{(cond)}}(T)
=
\frac{\left|
\sum_{j=1}^{N}\alpha_j(T)e^{-i\phi_j}
\right|^2}
{N\sum_{j=1}^{N}|\alpha_j(T)|^2}.
\label{eq:Fcond_perturbed_disc}
\end{equation}
For weak arm-dependent deviations, writing
\begin{align}
\alpha_j(T) &= \bar\alpha(T)\bigl[1+\varepsilon_j(T)\bigr],  \
|\varepsilon_j(T)|  \ll 1,
\label{eq:epsilon_def_disc}
\end{align}
one obtains
\begin{equation}
F_N^{\mathrm{(cond)}}(T)
\simeq
1-\frac{1}{N}\sum_{j=1}^{N}\left|\varepsilon_j-\bar\varepsilon\right|^2,\
\bar\varepsilon=\frac{1}{N}\sum_{j=1}^{N}\varepsilon_j.
\label{eq:Fcond_variance_disc}
\end{equation}
Thus, the leading reduction of the conditioned fidelity is controlled not by an overall rescaling of the transfer amplitudes, but by their {relative variance} across different arms.
This makes explicit why symmetry is the key ingredient protecting the ideal multipartite output.

For the tripartite protocol, two representative perturbation scans are particularly instructive.
A coupling-mismatch scan may be introduced as
\begin{align}
g_1  = g(1+\eta),  \
g_2  = g, \
g_3  = g(1-\eta),
\label{eq:g_mismatch_scan_disc}
\end{align}
while a detuning-mismatch scan may be introduced as
\begin{align}
\Delta_1  = \Delta+\delta_\Delta, \
\Delta_2  = \Delta,  \
\Delta_3  = \Delta-\delta_\Delta.
\label{eq:Delta_mismatch_scan_disc}
\end{align}
The first probes unequal local transfer strengths, whereas the second introduces relative phase mismatch between different transfer pathways.
In both cases, the most informative observables are the heralding probability, the conditional fidelity, the pathway-weight distribution
\begin{equation}
\frac{|\alpha_j(T)|^2}{\sum_{k=1}^{N}|\alpha_k(T)|^2},
\label{eq:alpha_weight_plot_disc}
\end{equation}
and the witness expectation value.

\begin{figure}[htpb]
\centering
\scalebox{0.23}{\includegraphics{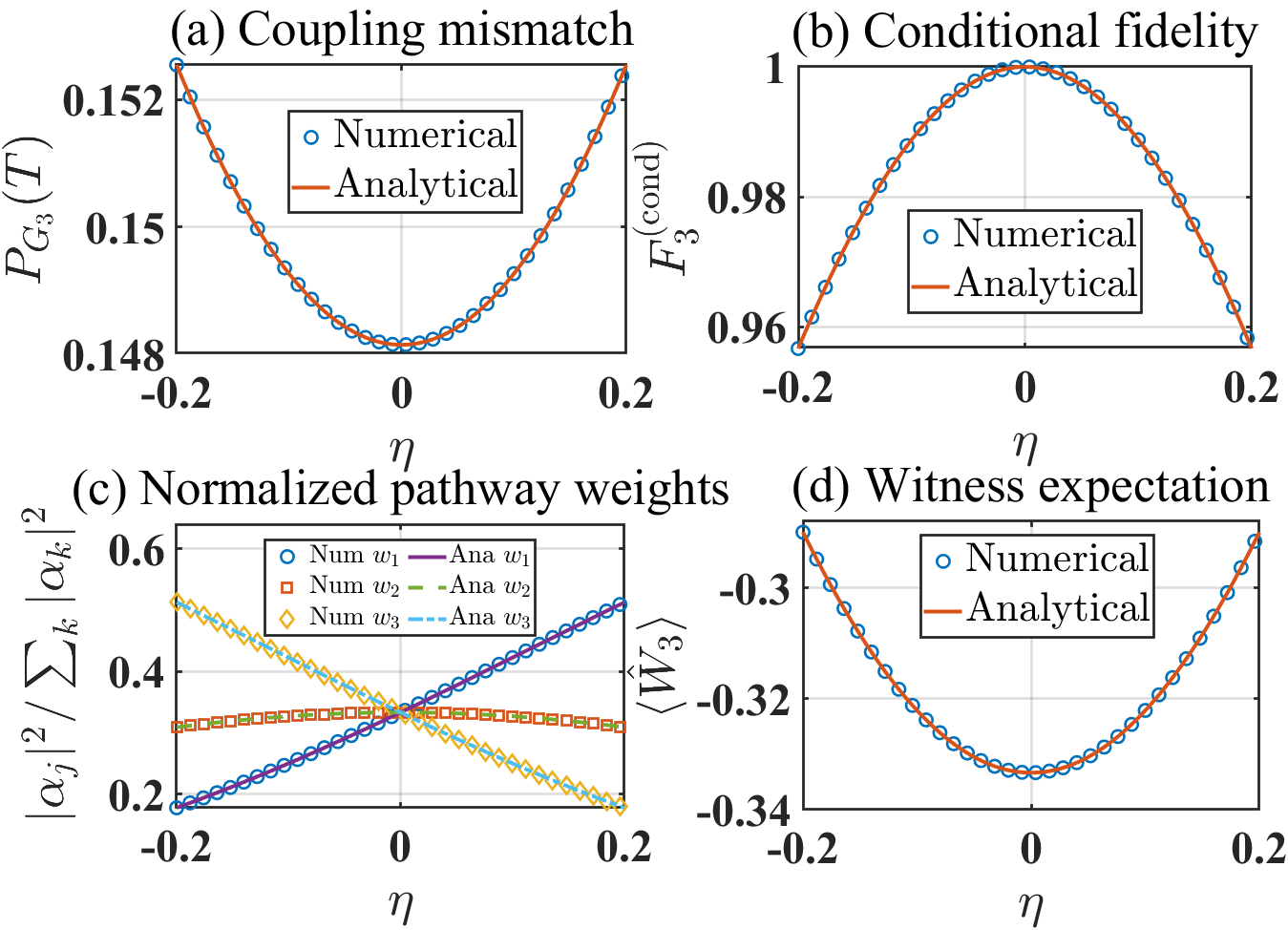}}
\caption{
Robustness to coupling-strength mismatch in the tripartite protocol for $N=3$.
A symmetric coupling imbalance is introduced according to Eq.~(\ref{eq:g_mismatch_scan_disc}).
(a) The heralding probability $P_{G_3}(T)$ as a function of the mismatch parameter $\eta$;
(b) The conditional fidelity $F^{(\mathrm{cond})}_3(T)$ defined by Eq.~(\ref{eq:Fcond_perturbed_disc}) as a function of $\eta$;
(c) The normalized pathway weights $|\alpha_j(T)|^2/\sum_{k=1}^{N}|\alpha_k(T)|^2$ defined by Eq.~(\ref{eq:alpha_weight_plot_disc}) as functions of $\eta$;
(d) The witness expectation value associated with the tripartite witness operator in Eq.~(\ref{eq:witness_3}) as a function of $\eta$.
Parameters: $g=\arccos\sqrt{2/3}$ and $\Delta=0$.
}
\label{fig:dis_coupling_mismatch}
\end{figure}

\begin{figure}[htpb]
\centering
\scalebox{0.23}{\includegraphics{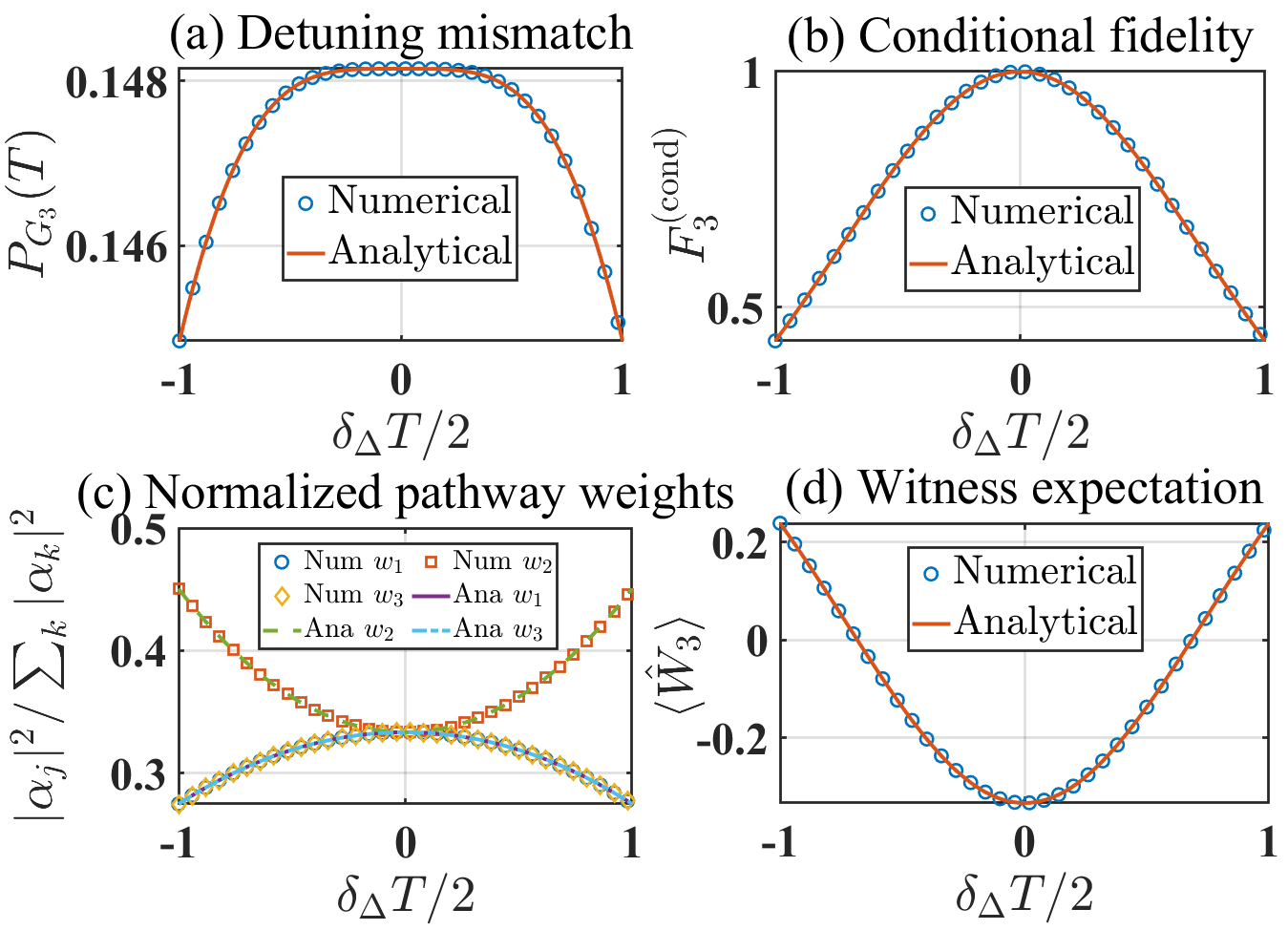}}
\caption{
Robustness to detuning mismatch in the tripartite protocol for $N=3$.
A symmetric detuning imbalance is introduced according to Eq.~(\ref{eq:Delta_mismatch_scan_disc}).
(a) The heralding probability $P_{G_3}(T)$ as a function of the normalized detuning mismatch $\delta_\Delta T/2$;
(b) The conditional fidelity $F^{(\mathrm{cond})}_3(T)$ defined by Eq.~(\ref{eq:Fcond_perturbed_disc}) as a function of $\delta_\Delta T/2$;
(c) The normalized pathway weights $|\alpha_j(T)|^2/\sum_{k=1}^{N}|\alpha_k(T)|^2$ defined by Eq.~(\ref{eq:alpha_weight_plot_disc}) as functions of $\delta_\Delta T/2$;
(d) The witness expectation value associated with the tripartite witness operator in Eq.~(\ref{eq:witness_3}) as a function of $\delta_\Delta T/2$.
Parameters: $g=\arccos\sqrt{2/3}$ and $\Delta=0$.
}
\label{fig:dis_detuning_mismatch}
\end{figure}

The numerical results confirm the analytical picture.
For coupling mismatch, Fig.~\ref{fig:dis_coupling_mismatch} shows that the heralding probability is only weakly affected over the scanned range, whereas the pathway weights become increasingly unbalanced as $|\eta|$ grows.
Correspondingly, the conditional fidelity decreases from unity and the witness expectation value moves upward toward zero, indicating a gradual degradation of the ideal tripartite structure.
For detuning mismatch, Fig.~\ref{fig:dis_detuning_mismatch} shows a similar separation between efficiency and structure, but with a stronger sensitivity of the conditional fidelity.
In that case, the dominant effect is not only amplitude imbalance but also detuning-induced relative phase accumulation, which modifies the interference between different heralded pathways.
As a result, detuning mismatch deforms the conditioned state more efficiently than common detuning, even when the overall success probability remains comparatively robust.

The results in  Figs.~\ref{fig:dis_coupling_mismatch} and ~\ref{fig:dis_detuning_mismatch} establish a clear hierarchy of robustness.
A common detuning mainly suppresses the heralding probability while leaving the postselected multipartite structure intact.
Weak symmetry breaking, by contrast, converts the ideal symmetric $W_3$ state into a weighted-$W_3$ state and thereby reduces the conditional fidelity and witness-based entanglement certification.
The protocol is therefore robust at the level of {transfer efficiency} against moderate imperfections, but the {quality} of the heralded multipartite state is protected specifically by symmetry.

\subsection{Beyond-RWA corrections and asymptotic validity}

The RWA  provides an analytically transparent description of the multipartite transfer mechanism.
A realistic treatment, however, must include the counter-rotating contributions present in the full interaction Hamiltonian.
The central question is therefore not the existence of beyond-RWA effects, but rather their structure and asymptotic scaling.
For each local arm, the leading effect of the fast counter-rotating terms is a Bloch--Siegert-type renormalization of the detuning.
The effective detuning is then given by
\begin{align}
\Delta &\rightarrow \Delta_{\mathrm{eff}}=\Delta+\Delta_{\mathrm{BS}},  \ \
\Delta_{\mathrm{BS}}  \simeq \kappa\frac{2G_0^2}{\Omega_+},
\label{eq:BS_shift_disc_new}
\end{align}
where $\kappa\sim 1$ is an envelope-dependent coefficient.
Physically, this shift originates from the residual dressing of the resonant sector by rapidly oscillating off-resonant harmonics.
Under the fixed-area condition $g=G_0T$, the Bloch--Siegert contribution scales as $\Delta_{\mathrm{BS}}  \propto {1}/{T^2}$.
Increasing the pulse duration therefore suppresses the instantaneous detuning shift.

An improved analytical description is obtained by replacing $\delta$ with the effective detuning parameter
\begin{equation}
\delta \rightarrow \delta_{\mathrm{eff}} = \frac{\Delta_{\mathrm{eff}} T}{2}
= \frac{\Delta T}{2} + \kappa\frac{g^2}{\Omega_+ T}.
\label{eq:delta_eff_disc_new}
\end{equation}
Substituting $\delta_{\mathrm{eff}}$ into the tripartite detuned RWA expression in Eq.~(\ref{eq:Pggg_detuned}) yields a Bloch--Siegert-corrected approximation for the heralding probability, which preserves the functional form of the RWA result while incorporating the dominant phase renormalization.

Beyond this deterministic shift, counter-rotating terms also populate off-resonant channels outside the target manifold.
For a square pulse, the leakage amplitude satisfies
\begin{equation}
\left|A_{\mathrm{leak}}\right|
\lesssim
\left|
\int_0^T dt\, G_0 e^{\pm i\Omega_+ t}
\right|,
\end{equation}
which leads to the estimate
\begin{equation}
P_{\mathrm{leak}} = O\!\left(\frac{G_0^2}{\Omega_+^2}\right).
\end{equation}
Using $g=G_0T$, one obtains
\begin{equation}
P_{\mathrm{leak}}
=
O\!\left(\frac{g^2}{\Omega_+^2 T^2}\right),
\label{eq:Pleak_T2_disc_new}
\end{equation}
revealing an algebraic suppression of beyond-RWA effects with increasing pulse duration.

\begin{figure}[htpb]
\centering
\scalebox{0.3}{\includegraphics{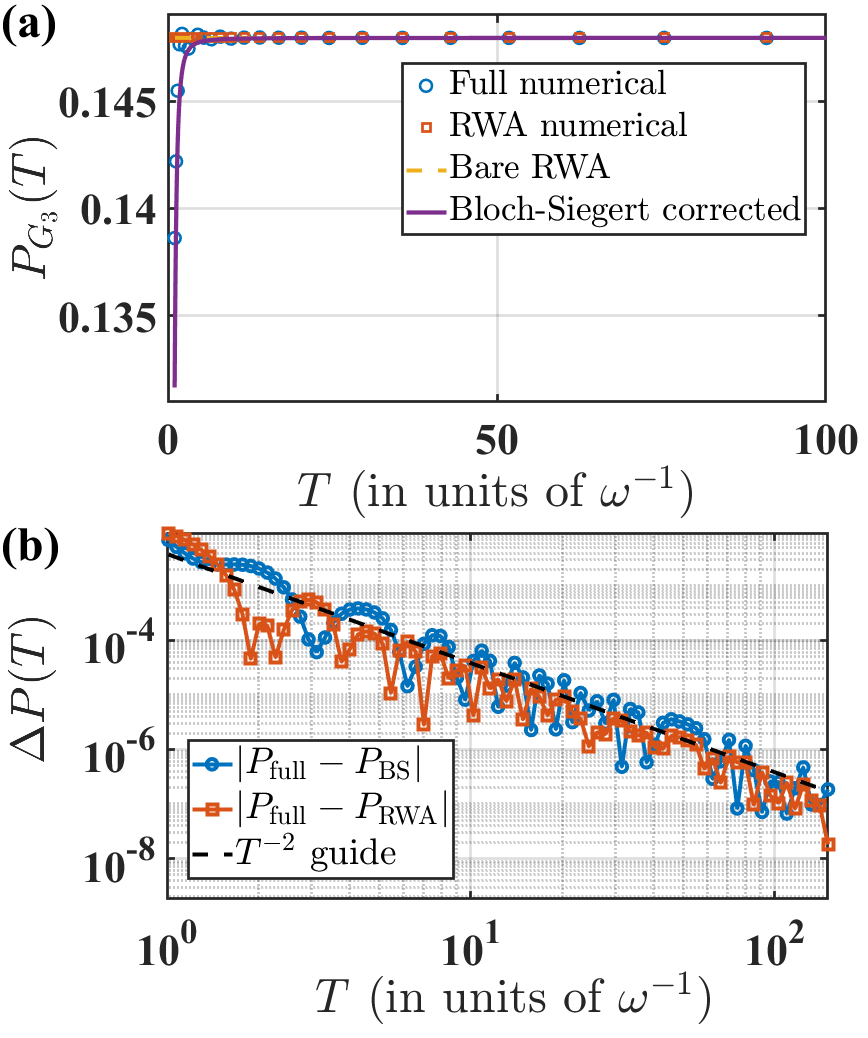}}
\caption{
Beyond-RWA corrections and asymptotic validity under a fixed pulse-area scan for the tripartite protocol with $N=3$.
The interaction duration $T$ is expressed in units of $\omega^{-1}$.
(a) The heralding probability $P_{G_3}(T)$ as a function of $T$, including the full numerical result obtained from the complete interaction Hamiltonian in Eq.~(\ref{eq:HI_full}), the RWA numerical result obtained from Eq.~(\ref{eq:HI_RWA}), the bare RWA analytical result from Eq.~(\ref{eq:Pggg_detuned}) at $\Delta=0$, and the Bloch--Siegert-corrected analytical result obtained from Eq.~(\ref{eq:Pggg_detuned}) under the substitution $\delta\rightarrow\delta_{\mathrm{eff}}$ in Eq.~(\ref{eq:delta_eff_disc_new});
(b) The residual discrepancies $\Delta P_{\mathrm{RWA}}(T)$ and $\Delta P_{\mathrm{BS}}(T)$ as functions of $T$ on a log-log scale, together with the $T^{-2}$ guide line associated with the asymptotic estimate in Eq.~(\ref{eq:Pleak_T2_disc_new}).
Parameters: $g=0.6$, $\Delta=0$ and $\kappa=6$.
}
\label{fig:dis_BS}
\end{figure}

Figure~\ref{fig:dis_BS}(a) compares the heralding probability $P_{G_3}(T)$ obtained from four descriptions of the tripartite dynamics under a fixed pulse-area scan.
The full numerical data are generated from the complete interaction Hamiltonian in Eq.~(\ref{eq:HI_full}), while the RWA numerical data are obtained from the reduced Hamiltonian in Eq.~(\ref{eq:HI_RWA}).
The bare RWA analytical curve is evaluated from Eq.~(\ref{eq:Pggg_detuned}) at $\Delta=0$, and the Bloch--Siegert-corrected curve is obtained from the same expression after the replacement $\delta\rightarrow\delta_{\mathrm{eff}}$ in Eq.~(\ref{eq:delta_eff_disc_new}).
Over the whole scan range, the RWA numerical data remain nearly indistinguishable from the bare RWA analytical prediction, which confirms the internal consistency of the RWA treatment.
Noticeable deviations arise only when the full dynamics is compared with the RWA-based descriptions.
For short interaction times, counter-rotating contributions reduce the heralding probability and shift the effective resonance condition away from the bare RWA prediction.
The Bloch--Siegert-corrected approximation reproduces that shift and follows the full numerical trend more closely in the short-pulse regime.
As $T$ increases, all curves approach the same asymptotic limit, showing that the beyond-RWA correction becomes progressively weaker for long pulses.

Figure~\ref{fig:dis_BS}(b) presents the residual discrepancies $\Delta P_{\mathrm{RWA}}(T)=|P_{\mathrm{full}}(T)-P_{\mathrm{RWA}}(T)|$ and $\Delta P_{\mathrm{BS}}(T)=|P_{\mathrm{full}}(T)-P_{\mathrm{BS}}(T)|$ on a log-log scale.
Both residuals decrease approximately as $T^{-2}$ in the asymptotic regime, in agreement with the leakage estimate in Eq.~(\ref{eq:Pleak_T2_disc_new}).
Such behavior indicates that the dominant beyond-RWA correction is algebraically suppressed when the pulse duration increases at fixed area.
The two residual curves remain close to one another throughout the scan, which shows that the Bloch--Siegert correction captures the leading resonance renormalization but does not remove the full algebraic discrepancy in the heralding probability.
The results support a simple physical picture.
For short pulses, beyond-RWA effects appear through resonance renormalization and leakage induced by counter-rotating channels.
For long pulses, rapid off-resonant oscillations are averaged out, and the multipartite transfer dynamics approaches the RWA limit in a controlled algebraic manner.

\subsection{Gaussian envelopes and pulse-shape effects on final electron entanglement}

Here we focus on the final unconditional electron entanglement to quantify how a smooth coupling profile changes the total transferred weight before heralding.
In realistic free-electron--TLS interactions, the coupling envelope is determined by the spatial variation of the near field sampled along the electron trajectory and is therefore naturally smooth rather than rectangular.
A convenient model is a Gaussian profile mapped from space to time,
\begin{equation}
G(t)=G_{\max}\exp\!\left[-\frac{(t-T/2)^2}{2\tau^2}\right],
\end{equation}
where $G_{\max}$ denotes the peak coupling strength and $\tau$ is the temporal width.
Unlike the equal-area normalization often adopted to isolate pure pulse-shape effects, the present discussion keeps the peak coupling fixed while varying $\tau$.
The accumulated interaction area then reads
\begin{equation}
g(\tau)=\int_0^T G(t)\,dt,
\end{equation}
so a change in the Gaussian width modifies not only the temporal profile but also the effective total interaction strength.

Such a distinction is physically relevant.
At exact resonance, the RWA dynamics generated by any real envelope is governed primarily by the accumulated area.
Under an equal-area normalization, Gaussian and square pulses therefore differ mainly through higher-order corrections and residual pulse-shape effects.
Under the fixed-peak normalization adopted here, however, increasing $\tau$ enlarges the actual interaction area $g(\tau)$ and thereby changes the amount of entanglement transferred to the electron subsystem.
This setting is closer to a realistic free-electron--TLS geometry, where the interaction is strongest near the closest approach and weakens away from that region.

\begin{figure}[htpb]
\centering
\scalebox{0.32}{\includegraphics{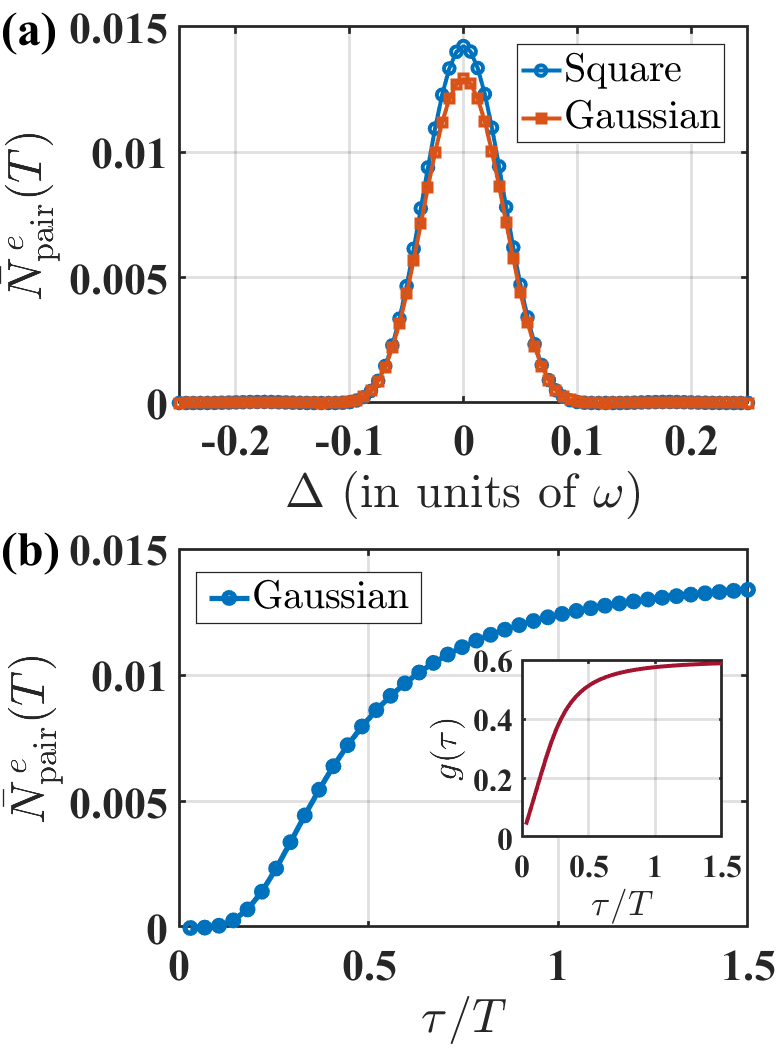}}
\caption{
Effects of fixed-peak Gaussian envelopes on the final unconditional electron entanglement in the tripartite protocol for $N=3$.
The interaction duration $T$ is expressed in units of $\omega^{-1}$.
The plotted quantity $\bar{\mathcal N}^{\,e}_{\mathrm{pair}}(T)$ denotes the average pairwise negativity of the final three-electron reduced state after tracing out the TLS degrees of freedom.
(a) $\bar{\mathcal N}^{\,e}_{\mathrm{pair}}(T)$ as a function of the common detuning $\Delta$, comparing a square pulse and a Gaussian envelope ($\tau/T=1.2$) with the same peak coupling strength;
(b) $\bar{\mathcal N}^{\,e}_{\mathrm{pair}}(T)$ as a function of the relative Gaussian width $\tau/T$ at fixed peak coupling, fixed interaction window $T$, and $\Delta=0$.
The inset in (b) shows the accumulated pulse area $g(\tau)=\int_0^T G(t)\,dt$ as a function of $\tau/T$.
Parameters: $G_{\max}=G_0$.
}
\label{fig:dis_Gaus}
\end{figure}

To characterize the final multipartite transfer performance, we consider the unconditional three-electron output state obtained after tracing out the TLS degrees of freedom and use the average pairwise negativity
\begin{equation}
\bar{\mathcal N}^{\,e}_{\mathrm{pair}}(T)
=
\frac{
\mathcal N^{e}_{12}(T)+\mathcal N^{e}_{13}(T)+\mathcal N^{e}_{23}(T)
}{3}
\end{equation}
as the entanglement measure.
Figure~\ref{fig:dis_Gaus} summarizes how the pulse shape influences the final unconditional electron entanglement.
Figure~\ref{fig:dis_Gaus}(a) presents $\bar{\mathcal N}^{\,e}_{\mathrm{pair}}(T)$ as a function of the common detuning $\Delta$ for a square pulse and for a Gaussian envelope with the same peak coupling strength.
Both curves retain the same overall resonant structure.
The final electron entanglement is maximal near $\Delta=0$ and decreases as the system moves away from resonance.
The Gaussian curve remains slightly below the square-pulse curve in the vicinity of resonance.
Such behavior is expected under fixed-peak normalization, because the Gaussian envelope spends less time near the maximum coupling strength and therefore accumulates a smaller effective interaction area than the square pulse over the same finite time window.

Figure~\ref{fig:dis_Gaus}(b) shows $\bar{\mathcal N}^{\,e}_{\mathrm{pair}}(T)$ at $\Delta=0$ as a function of the relative Gaussian width $\tau/T$.
The final unconditional electron entanglement increases monotonically with $\tau/T$ and gradually approaches the square-pulse limit for broad Gaussian envelopes.
The inset displays the accumulated interaction area $g(\tau)$, which follows the same trend and saturates as $\tau/T$ becomes large.
The close correspondence between the main panel and the inset indicates that, within the fixed-peak setting, the dominant role of increasing the Gaussian width is to enlarge the effective interaction area rather than to alter the underlying transfer mechanism.

The results in Figs.~\ref{fig:dis_Gaus}(a) and (b) clarify the influence of realistic smooth envelopes on multipartite entanglement transfer.
Under fixed-peak driving, a Gaussian profile acts mainly through the growth of the accumulated coupling area, which enhances the final unconditional electron entanglement and gradually brings the dynamics closer to the square-pulse limit.
At the same time, the resonant structure and the single-excitation transfer mechanism remain unchanged.
Gaussian envelopes therefore provide a realistic extension of the square-pulse model without modifying the central physical picture of the protocol.

\subsection{Physical implications, limitations, and outlook}

The present multipartite protocol significantly extends the previous work on heralded entanglement transfer between an atomic pair and two free electrons \cite{ran2026heralded}.
A fundamental distinction lies in the structure of the generated electron states: the two-body protocol allows for the heralded preparation of entangled states in both the $\pm 1$ sideband manifolds, effectively supporting two distinct Bell-state branches.
In contrast, the multipartite protocol is constrained to the generation of a single $|W_N^{(+)}\rangle$ state in the upper-sideband manifold.
This difference arises because the heralded multipartite branch requires the simultaneous survival of all $N-1$ spectator arms in the ground-sideband sector, which strictly selects the upper-sideband excitation channel and suppresses the lower-sideband leakage.
Despite this difference in manifold accessibility, both protocols share a common reliance on the preservation of initial atomic phase coherence.
In both the two-body and multipartite cases, the relative phases $\phi_j$ of the initial atomic resource are mapped directly onto the electronic state, ensuring that the heralded electron entanglement inherits the internal phase structure of the atomic input.
This phase-preserving property confirms that the heralded transfer is an isometric mapping, where the multipartite entanglement structure is protected by the symmetry of the local interaction arms, regardless of the system size $N$.

Our results identify a concrete route for generating multipartite free-electron entanglement from strictly local electron--TLS interactions.
The essential mechanism combines an initially prepared multipartite atomic coherence with a heralding measurement that removes which-arm information, rather than relying on any direct nonlocal interaction between different electrons.
The transfer process thereby maps a localized single-excitation manifold of matter qubits onto a delocalized single-excitation manifold of propagating electrons.
For quantum electron optics, the protocol provides a transparent mechanism by which multipartite electron entanglement can emerge from sideband-resolved local exchange and postselected interference among indistinguishable transfer pathways.
An experimental implementation requires a spatially separated array of localized TLSs prepared in a single-excitation $W_N$ resource, synchronized free-electron trajectories coupled predominantly to designated local targets, and a heralding stage capable of resolving the all-ground TLS outcome together with electron-sideband readout.
Within such a platform, the symmetric configuration analyzed here defines the clean operating point, while detuning, coupling asymmetry, pulse shape, and beyond-RWA corrections provide practical tolerance criteria.
Recent experimental demonstrations of free-electron--photon entanglement and coincidence-based quantum imaging in transmission electron microscopy platforms further support the feasibility of coherent quantum-correlation protocols involving propagating electron wave packets \cite{henke2025observation,preimesberger2025experimental}.
In contrast to these hybrid electron--photon entangled states, the present protocol addresses the heralded generation of multipartite entanglement directly between free electrons.

Limitations:
The protocol is probabilistic, and even under optimal resonant conditions the heralding probability decreases algebraically with system size because one desired transfer event must coexist with survival of all spectator arms.
The exact $W_N$ structure also relies on symmetry, since arm-dependent perturbations deform the heralded output into a weighted single-excitation state.
The analytical treatment is mainly based on the rotating-wave approximation and independent local interaction arms.
Although the numerical results show that beyond-RWA corrections remain perturbative in the appropriate regime, short pulses or stronger couplings can produce measurable deviations through counter-rotating channels.
The present theory is restricted to the single-excitation sector and does not yet include multiple-excitation resources, higher sideband occupancies, mixed initial states, measurement imperfections, decoherence, or partial distinguishability between interaction arms.
For larger systems, pairwise negativities become progressively less informative, making manifold-resolved fidelities and genuine multipartite witnesses more suitable observables.

Outlook: Weighted single-excitation resources can be treated as programmable inputs rather than only as symmetry-breaking imperfections, enabling tailored multipartite electron-state preparation.
More realistic models should incorporate smooth spatial envelopes, electron wave-packet effects, imperfect TLS readout, finite decoherence, and partial distinguishability among local transfer pathways.
Richer atomic inputs beyond the single-excitation sector may also be mapped onto higher-dimensional electron-sideband manifolds with multipartite structures beyond the $W$ class.
On the experimental side, the protocol suggests hybrid quantum electron-optical architectures in which localized matter systems act as entanglement resources and free electrons serve as mobile carriers of nonclassical correlations.
Related proposals based on entanglement swapping and quantum-eraser protocols have further suggested that electron-mediated quantum correlations may enable quantum-enhanced nanoscale probing and spectroscopy \cite{henke2025probing}.
Such capabilities would be relevant for multipartite entanglement in propagating massive particles, quantum-enabled electron interferometry, heralded state engineering, and the transfer of nonclassical correlations between stationary and flying quantum platforms.
The main contribution of the present work is therefore a minimal and scalable physical template for heralded multipartite entanglement transfer in QEO.

\section{Conclusion}\label{CON}
We have studied heralded multipartite entanglement transfer from atomic single-excitation $W_N$ resources to free electrons in a sideband-resolved interaction picture.
For a symmetric $N$-arm electron--TLS geometry, the heralded branch was derived analytically within the RWA.
Postselection of the all-ground TLS outcome maps the atomic single-excitation manifold onto the electronic upper-sideband single-excitation manifold and yields an exact $N$-electron $W_N$-type entangled state.
The heralding probability was obtained in closed form for resonant and detuned regimes, with the resonant optimum scaling as $P_{G_N}^{\max}\sim e^{-1}/N$ at large $N$.
The entanglement structure was analyzed for arbitrary $N$ and illustrated explicitly for $N=3$, where correlation redistribution and weighted single-excitation transfer were examined in detail.
Common detuning mainly reduces the heralding probability, whereas weak symmetry breaking converts the ideal symmetric output into a weighted single-excitation state.
Beyond-RWA effects are algebraically suppressed for long pulses, and Gaussian envelopes preserve the same transfer mechanism while modifying the effective interaction strength under fixed-peak driving.
The main contribution of the present work is the identification of a minimal heralded mechanism by which multipartite atomic coherence can be converted into multipartite free-electron entanglement through local electron--TLS exchange.

\
\

\noindent {\bf Acknowledgments}
This work is supported by the Natural Science Foundation of Chongqing (Grant No. CSTB2025NSCQ-GPX0416) and the Science and Technology Research Program of Chongqing Municipal Education Commission (Grant No. KJQN202401437).
S. L. acknowledges the support of the Natural Science Foundation of Hubei Province (Grant No. 2024AFB200), Teacher Research Ability Cultivation Fund of Hubei University of Arts and Sciences (Grant No. 2023pygpzk05), and the Research Start-up Fund of Hubei University of Arts and Sciences (Grant No.
qdf2022033).
Y.-D. L acknowledges the Support of the Science and Technology Research Program of Chongqing Municipal Education Commission (Grant No. KJZD-K202401403) and the scientific research project of the Science and Technology Bureau of Fuling (Grant No. FLKJ2025AAG2003).
A. G. and R. I. acknowledge the support of the Israel Science Foundation (Grant No. 2992124).

\bibliographystyle{unsrt}
\bibliography{references}

@article{ran2026heralded,
  title={Heralded Entanglement Transfer from Entangled Atomic Pair to Free Electrons},
  author={Ran, Du and Ianconescu, Reuven and Liu, Shuai and Li, Ya-dong and Bai, Ji-Yuan and He, Ze-Long and Shi, Zhi-Cheng and Xia, Yan and Gover, Avraham},
  journal={arXiv preprint arXiv:2604.22974},
  year={2026}
}

@article{zhang2022quantum,
  title={Quantum state interrogation using a preshaped free electron wavefunction},
  author={Zhang, Bin and Ran, Du and Ianconescu, Reuven and Friedman, Aharon and Scheuer, Jacob and Yariv, Amnon and Gover, Avraham},
  journal={Physical Review Research},
  volume={4},
  number={3},
  pages={033071},
  year={2022},
  publisher={APS}
}

@article{bucher2023free,
  title={Free-electron Ramsey-type interferometry for enhanced amplitude and phase imaging of nearfields},
  author={Bucher, Tomer and Ruimy, Ron and Tsesses, Shai and Dahan, Raphael and Bartal, Guy and Vanacore, Giovanni Maria and Kaminer, Ido},
  journal={Science Advances},
  volume={9},
  number={51},
  pages={eadi5729},
  year={2023},
  publisher={American Association for the Advancement of Science}
}

@article{tsesses2023tunable,
  title={Tunable photon-induced spatial modulation of free electrons},
  author={Tsesses, Shai and Dahan, Raphael and Wang, Kangpeng and Bucher, Tomer and Cohen, Kobi and Reinhardt, Ori and Bartal, Guy and Kaminer, Ido},
  journal={Nature Materials},
  volume={22},
  number={3},
  pages={345--352},
  year={2023},
  publisher={Nature Publishing Group UK London}
}

@article{luo2019quantum,
  title={Quantum teleportation in high dimensions},
  author={Luo, Yi-Han and Zhong, Han-Sen and Erhard, Manuel and Wang, Xi-Lin and Peng, Li-Chao and Krenn, Mario and Jiang, Xiao and Li, Li and Liu, Nai-Le and Lu, Chao-Yang and others},
  journal={Physical review letters},
  volume={123},
  number={7},
  pages={070505},
  year={2019},
  publisher={APS}
}

@article{ekert1991quantum,
  title={Quantum cryptography based on Bell’s theorem},
  author={Ekert, Artur K},
  journal={Physical review letters},
  volume={67},
  number={6},
  pages={661},
  year={1991},
  publisher={APS}
}

@article{bennett1993teleporting,
  title={Teleporting an unknown quantum state via dual classical and Einstein-Podolsky-Rosen channels},
  author={Bennett, Charles H and Brassard, Gilles and Cr{\'e}peau, Claude and Jozsa, Richard and Peres, Asher and Wootters, William K},
  journal={Physical review letters},
  volume={70},
  number={13},
  pages={1895},
  year={1993},
  publisher={APS}
}

@article{gottesman1999demonstrating,
  title={Demonstrating the viability of universal quantum computation using teleportation and single-qubit operations},
  author={Gottesman, Daniel and Chuang, Isaac L},
  journal={Nature},
  volume={402},
  number={6760},
  pages={390--393},
  year={1999},
  publisher={Nature Publishing Group UK London}
}

@article{paneru2021experimental,
  title={Experimental tests of multiplicative Bell inequalities and the fundamental role of local correlations},
  author={Paneru, Dilip and Te'eni, Amit and Peled, Bar Y and Hubble, James and Zhang, Yingwen and Carmi, Avishy and Cohen, Eliahu and Karimi, Ebrahim},
  journal={Physical Review Research},
  volume={3},
  number={1},
  pages={L012025},
  year={2021},
  publisher={APS}
}

@article{aspect1982experimental,
  title={Experimental test of Bell's inequalities using time-varying analyzers},
  author={Aspect, Alain and Dalibard, Jean and Roger, G{\'e}rard},
  journal={Physical review letters},
  volume={49},
  number={25},
  pages={1804},
  year={1982},
  publisher={APS}
}

@article{forbes2025heralded,
  title={Heralded generation of entanglement with photons},
  author={Forbes, Imogen and Ghafari, Farzad and Deacon, Edward CR and Singh, Sukhjit P and Lavie, Emilien and Yard, Patrick and Shaw, Reece D and Laing, Anthony and Tischler, Nora},
  journal={Reports on Progress in Physics},
  volume={88},
  number={8},
  pages={086002},
  year={2025},
  publisher={IOP Publishing}
}

@article{chen2021quantum,
  title={Quantum entanglement on photonic chips: a review},
  author={Chen, Xiaojiong and Fu, Zhaorong and Gong, Qihuang and Wang, Jianwei},
  journal={Advanced Photonics},
  volume={3},
  number={6},
  pages={064002--064002},
  year={2021},
  publisher={Society of Photo-Optical Instrumentation Engineers}
}

@article{wilk2010entanglement,
  title={Entanglement of two individual neutral atoms using Rydberg blockade},
  author={Wilk, Tatjana and Ga{\"e}tan, A and Evellin, C and Wolters, J and Miroshnychenko, Y and Grangier, P and Browaeys, Antoine},
  journal={Physical review letters},
  volume={104},
  number={1},
  pages={010502},
  year={2010},
  publisher={APS}
}

@article{shao2017ground,
  title={Ground-state blockade of Rydberg atoms and application in entanglement generation},
  author={Shao, XQ and Li, DX and Ji, YQ and Wu, JH and Yi, XX},
  journal={Physical Review A},
  volume={96},
  number={1},
  pages={012328},
  year={2017},
  publisher={APS}
}

@article{chen2025hardware,
  title={Hardware-efficient stabilization of entanglement via engineered dissipation in superconducting circuits},
  author={Chen, Changling and Tang, Kai and Zhou, Yuxuan and Yi, KangYuan and Zhang, Xuan and Zhang, Xu and Guo, Haosheng and Liu, Song and Chen, Yuanzhen and Yan, Tongxing and others},
  journal={Physical Review Research},
  volume={7},
  number={2},
  pages={L022018},
  year={2025},
  publisher={APS}
}

@article{zhang2023generating,
  title={Generating Bell states and N-partite W states of long-distance qubits in superconducting waveguide QED},
  author={Zhang, Guo-Qiang and Feng, Wei and Xiong, Wei and Xu, Da and Su, Qi-Ping and Yang, Chui-Ping},
  journal={Physical Review Applied},
  volume={20},
  number={4},
  pages={044014},
  year={2023},
  publisher={APS}
}

@article{zheng2000efficient,
  title={Efficient scheme for two-atom entanglement and quantum information processing in cavity QED},
  author={Zheng, Shi-Biao and Guo, Guang-Can},
  journal={Physical Review Letters},
  volume={85},
  number={11},
  pages={2392},
  year={2000},
  publisher={APS}
}

@article{zheng2001one,
  title={One-step synthesis of multiatom Greenberger-Horne-Zeilinger states},
  author={Zheng, Shi-Biao},
  journal={Physical review letters},
  volume={87},
  number={23},
  pages={230404},
  year={2001},
  publisher={APS}
}

@article{yuan2020steady,
  title={Steady bell state generation via magnon-photon coupling},
  author={Yuan, HY and Yan, Peng and Zheng, Shasha and He, QY and Xia, Ke and Yung, Man-Hong},
  journal={Physical Review Letters},
  volume={124},
  number={5},
  pages={053602},
  year={2020},
  publisher={APS}
}

@article{zou2022bell,
  title={Bell-state generation for spin qubits via dissipative coupling},
  author={Zou, Ji and Zhang, Shu and Tserkovnyak, Yaroslav},
  journal={Physical Review B},
  volume={106},
  number={18},
  pages={L180406},
  year={2022},
  publisher={APS}
}

@article{weng2025high,
  title={High-fidelity generation of Bell and W states in a giant-atom system via bound states in the continuum},
  author={Weng, Mingzhu and Yu, Hongwei and Wang, Zhihai},
  journal={Physical Review A},
  volume={111},
  number={5},
  pages={053711},
  year={2025},
  publisher={APS}
}

@article{yang2018deterministic,
  title={Deterministic generation of Greenberger--Horne--Zeilinger entangled states of cat-state qubits in circuit QED},
  author={Yang, Chui-Ping and Zheng, Zhen-Fei},
  journal={Optics Letters},
  volume={43},
  number={20},
  pages={5126--5129},
  year={2018},
  publisher={Optical Society of America}
}

@article{zhang2024fast,
  title={Fast generation of GHZ-like states using collective-spin XYZ model},
  author={Zhang, Xuanchen and Hu, Zhiyao and Liu, Yong-Chun},
  journal={Physical Review Letters},
  volume={132},
  number={11},
  pages={113402},
  year={2024},
  publisher={APS}
}

@article{peng2021one,
  title={One-photon solutions to the multiqubit multimode quantum Rabi model for fast W-state generation},
  author={Peng, Jie and Zheng, Juncong and Yu, Jing and Tang, Pinghua and Barrios, G Alvarado and Zhong, Jianxin and Solano, Enrique and Albarr{\'a}n-Arriagada, F and Lamata, Lucas},
  journal={Physical Review Letters},
  volume={127},
  number={4},
  pages={043604},
  year={2021},
  publisher={APS}
}

@article{francesco2024steady,
  title={Steady-state entanglement production in a quantum thermal machine with continuous feedback control},
  author={Francesco Diotallevi, Giovanni and Annby-Andersson, Bj{\"o}rn and Samuelsson, Peter and Tavakoli, Armin and Bakhshinezhad, Pharnam},
  journal={New Journal of Physics},
  volume={26},
  number={5},
  pages={053005},
  year={2024},
  publisher={IOP Publishing}
}

@article{hashim2025efficient,
  title={Efficient generation of multi-partite entanglement between non-local superconducting qubits using classical feedback},
  author={Hashim, Akel and Yuan, Ming and Gokhale, Pranav and Chen, Larry and J{\"u}nger, Christian and Fruitwala, Neelay and Xu, Yilun and Huang, Gang and Nowrouzi, Kasra and Jiang, Liang and others},
  journal={APL Quantum},
  volume={2},
  number={4},
  year={2025},
  publisher={AIP Publishing}
}

@article{kastoryano2011dissipative,
  title={Dissipative preparation of entanglement in optical cavities},
  author={Kastoryano, Michael James and Reiter, Florentin and S{\o}rensen, Anders S{\o}ndberg},
  journal={Physical review letters},
  volume={106},
  number={9},
  pages={090502},
  year={2011},
  publisher={APS}
}

@article{zhang2023generation,
  title={Generation of long-lived W states via reservoir engineering in dissipatively coupled systems},
  author={Zhang, Guo-Qiang and Feng, Wei and Xiong, Wei and Su, Qi-Ping and Yang, Chui-Ping},
  journal={Physical Review A},
  volume={107},
  number={1},
  pages={012410},
  year={2023},
  publisher={APS}
}

@article{cole2022resource,
  title={Resource-efficient dissipative entanglement of two trapped-ion qubits},
  author={Cole, Daniel C and Erickson, Stephen D and Zarantonello, Giorgio and Horn, Karl P and Hou, Pan-Yu and Wu, Jenny J and Slichter, Daniel H and Reiter, Florentin and Koch, Christiane P and Leibfried, Dietrich},
  journal={Physical Review Letters},
  volume={128},
  number={8},
  pages={080502},
  year={2022},
  publisher={APS}
}

@article{xu2024efficient,
  title={Efficient generation of multiqubit entanglement states using rapid adiabatic passage},
  author={Xu, Shijie and Li, Xinwei and Li, Xiangliang and Li, Jinbin and Xue, Ming},
  journal={Physical Review A},
  volume={110},
  number={2},
  pages={023108},
  year={2024},
  publisher={APS}
}

@article{chang2020remote,
  title={Remote entanglement via adiabatic passage using a tunably dissipative quantum communication system},
  author={Chang, H-S and Zhong, YP and Bienfait, Audrey and Chou, M-H and Conner, Christopher R and Dumur, {\'E}tienne and Grebel, Joel and Peairs, Gregory A and Povey, Rhys G and Satzinger, Kevin J and others},
  journal={Physical Review Letters},
  volume={124},
  number={24},
  pages={240502},
  year={2020},
  publisher={APS}
}

@article{carrasco2024dicke,
  title={Dicke state generation and extreme spin squeezing via rapid adiabatic passage},
  author={Carrasco, Sebastian C and Goerz, Michael H and Malinovskaya, Svetlana A and Vuleti{\'c}, Vladan and Schleich, Wolfgang P and Malinovsky, Vladimir S},
  journal={Physical Review Letters},
  volume={132},
  number={15},
  pages={153603},
  year={2024},
  publisher={APS}
}

@article{barontini2015deterministic,
  title={Deterministic generation of multiparticle entanglement by quantum Zeno dynamics},
  author={Barontini, Giovanni and Hohmann, Leander and Haas, Florian and Est{\`e}ve, J{\'e}r{\^o}me and Reichel, Jakob},
  journal={Science},
  volume={349},
  number={6254},
  pages={1317--1321},
  year={2015},
  publisher={American Association for the Advancement of Science}
}

@article{liang2015adiabatic,
  title={Adiabatic passage for three-dimensional entanglement generation through quantum Zeno dynamics},
  author={Liang, Yan and Su, Shi-Lei and Wu, Qi-Cheng and Ji, Xin and Zhang, Shou},
  journal={Optics Express},
  volume={23},
  number={4},
  pages={5064--5077},
  year={2015},
  publisher={Optical Society of America}
}

@article{wang2008quantum,
  title={Quantum entanglement via two-qubit quantum Zeno dynamics},
  author={Wang, Xiang-Bin and You, JQ and Nori, Franco},
  journal={Physical Review A},
  volume={77},
  number={6},
  pages={062339},
  year={2008},
  publisher={APS}
}

@article{paternostro2004complete,
  title={Complete conditions for entanglement transfer},
  author={Paternostro, Mauro and Son, Wonmin and Kim, MS},
  journal={Physical review letters},
  volume={92},
  number={19},
  pages={197901},
  year={2004},
  publisher={APS}
}

@article{lee2006entanglement,
  title={Entanglement reciprocation between qubits and continuous variables},
  author={Lee, Jinhyoung and Paternostro, Mauro and Kim, MS and Bose, Sougato},
  journal={Physical review letters},
  volume={96},
  number={8},
  pages={080501},
  year={2006},
  publisher={APS}
}

@article{serafini2006enhanced,
  title={Enhanced dynamical entanglement transfer with multiple qubits},
  author={Serafini, A and Paternostro, Mauro and Kim, MS and Bose, S},
  journal={Physical Review A},
  volume={73},
  number={2},
  pages={022312},
  year={2006},
  publisher={APS}
}

@article{zou2006entanglement,
  title={Entanglement transfer from entangled two-mode fields to a pair of separable and mixed qubits},
  author={Zou, Jian and Li, Jun Gang and Shao, Bin and Li, Jian and ShuLi, Qian},
  journal={Physical Review A},
  volume={73},
  number={4},
  pages={042319},
  year={2006},
  publisher={APS}
}

@article{casagrande2007improving,
  title={Improving the entanglement transfer from continuous-variable systems to localized qubits using non-Gaussian states},
  author={Casagrande, Federico and Lulli, Alfredo and Paris, Matteo GA},
  journal={Physical Review A—Atomic, Molecular, and Optical Physics},
  volume={75},
  number={3},
  pages={032336},
  year={2007},
  publisher={APS}
}

@article{reinhardt2020theory,
  title={Theory of shaping electron wavepackets with light},
  author={Reinhardt, Ori and Kaminer, Ido},
  journal={ACS Photonics},
  volume={7},
  number={10},
  pages={2859--2870},
  year={2020},
  publisher={ACS Publications}
}

@article{madan2022ultrafast,
  title={Ultrafast transverse modulation of free electrons by interaction with shaped optical fields},
  author={Madan, Ivan and Leccese, Veronica and Mazur, Adam and Barantani, Francesco and LaGrange, Thomas and Sapozhnik, Alexey and Tengdin, Phoebe M and Gargiulo, Simone and Rotunno, Enzo and Olaya, Jean-Christophe and others},
  journal={ACS photonics},
  volume={9},
  number={10},
  pages={3215--3224},
  year={2022},
  publisher={ACS Publications}
}

@article{vanacore2018attosecond,
  title={Attosecond coherent control of free-electron wave functions using semi-infinite light fields},
  author={Vanacore, Giovanni M and Madan, I and Berruto, G and Wang, K and Pomarico, E and Lamb, RJ and McGrouther, D and Kaminer, I and Barwick, B and Garc{\'\i}a de Abajo, F Javier and others},
  journal={Nature communications},
  volume={9},
  number={1},
  pages={2694},
  year={2018},
  publisher={Nature Publishing Group UK London}
}

@article{wong2021control,
  title={Control of quantum electrodynamical processes by shaping electron wavepackets},
  author={Wong, Liang Jie and Rivera, Nicholas and Murdia, Chitraang and Christensen, Thomas and Joannopoulos, John D and Solja{\v{c}}i{\'c}, Marin and Kaminer, Ido},
  journal={Nature communications},
  volume={12},
  number={1},
  pages={1700},
  year={2021},
  publisher={Nature Publishing Group UK London}
}

@article{vanacore2020spatio,
  title={Spatio-temporal shaping of a free-electron wave function via coherent light--electron interaction},
  author={Vanacore, Giovanni Maria and Madan, Ivan and Carbone, Fabrizio},
  journal={La Rivista del Nuovo Cimento},
  volume={43},
  number={11},
  pages={567--597},
  year={2020},
  publisher={Springer}
}

@article{di2019probing,
  title={Probing quantum optical excitations with fast electrons},
  author={Di Giulio, Valerio and Kociak, Mathieu and de Abajo, F Javier Garc{\'\i}a},
  journal={Optica},
  volume={6},
  number={12},
  pages={1524--1534},
  year={2019},
  publisher={Optical Society of America}
}

@article{kfir2021optical,
  title={Optical coherence transfer mediated by free electrons},
  author={Kfir, Ofer and Di Giulio, Valerio and de Abajo, F Javier Garc{\'\i}a and Ropers, Claus},
  journal={Science Advances},
  volume={7},
  number={18},
  pages={eabf6380},
  year={2021},
  publisher={American Association for the Advancement of Science}
}

@article{sun2023generating,
  title={Generating optical cat states via quantum interference of multi-path free-electron--photons interactions},
  author={Sun, Feng-Xiao and Fang, Yiqi and He, Qiongyi and Liu, Yunquan},
  journal={Science Bulletin},
  volume={68},
  number={13},
  pages={1366--1371},
  year={2023},
  publisher={Elsevier}
}

@article{ben2021shaping,
  title={Shaping quantum photonic states using free electrons},
  author={Ben Hayun, Adi and Reinhardt, Ori and Nemirovsky, Jonathan and Karnieli, Aviv and Rivera, Nicholas and Kaminer, Ido},
  journal={Science Advances},
  volume={7},
  number={11},
  pages={eabe4270},
  year={2021},
  publisher={American Association for the Advancement of Science}
}

@article{tsarev2021measurement,
  title={Measurement of temporal coherence of free electrons by time-domain electron interferometry},
  author={Tsarev, M and Ryabov, Andrey and Baum, Peter},
  journal={Physical Review Letters},
  volume={127},
  number={16},
  pages={165501},
  year={2021},
  publisher={APS}
}

@article{garcia2021optical,
  title={Optical excitations with electron beams: challenges and opportunities},
  author={Garcia de Abajo, F Javier and Di Giulio, Valerio},
  journal={ACS photonics},
  volume={8},
  number={4},
  pages={945--974},
  year={2021},
  publisher={ACS Publications}
}

@article{arend2025electrons,
  title={Electrons herald non-classical light},
  author={Arend, Germaine and Huang, Guanhao and Feist, Armin and Yang, Yujia and Henke, Jan-Wilke and Qiu, Zheru and Jeng, Hao and Raja, Arslan Sajid and Haindl, Rudolf and Wang, Rui Ning and others},
  journal={Nature Physics},
  pages={1--8},
  year={2025},
  publisher={Nature Publishing Group UK London}
}

@article{gover2020free,
  title={Free-electron--bound-electron resonant interaction},
  author={Gover, Avraham and Yariv, Amnon},
  journal={Physical Review Letters},
  volume={124},
  number={6},
  pages={064801},
  year={2020},
  publisher={APS}
}

@article{zhang2021quantum,
  title={Quantum wave-particle duality in free-electron--bound-electron interaction},
  author={Zhang, Bin and Ran, Du and Ianconescu, Reuven and Friedman, Aharon and Scheuer, Jacob and Yariv, Amnon and Gover, Avraham},
  journal={Physical Review Letters},
  volume={126},
  number={24},
  pages={244801},
  year={2021},
  publisher={APS}
}

@article{ran2022coherent,
  title={Coherent excitation of bound electron quantum state with quantum electron wavepackets},
  author={Ran, Du and Zhang, Bin and Ianconescu, Reuven and Friedman, Aharon and Scheuer, Jacob and Yariv, Amnon and Gover, Avraham},
  journal={Frontiers in Physics},
  volume={10},
  pages={920701},
  year={2022},
  publisher={Frontiers Media SA}
}

@article{zhao2021quantum,
  title={Quantum entanglement and modulation enhancement of free-electron--bound-electron interaction},
  author={Zhao, Zhexin and Sun, Xiao-Qi and Fan, Shanhui},
  journal={Physical Review Letters},
  volume={126},
  number={23},
  pages={233402},
  year={2021},
  publisher={APS}
}

@article{ruimy2021toward,
  title={Toward atomic-resolution quantum measurements with coherently shaped free electrons},
  author={Ruimy, Ron and Gorlach, Alexey and Mechel, Chen and Rivera, Nicholas and Kaminer, Ido},
  journal={Physical Review Letters},
  volume={126},
  number={23},
  pages={233403},
  year={2021},
  publisher={APS}
}

@article{crispin2025probing,
  title={Probing quantum-coherent dynamics with free electrons},
  author={Crispin, HB and Talebi, N},
  journal={arXiv preprint arXiv:2512.24883},
  year={2025}
}

@article{yalunin2021tailored,
  title={Tailored high-contrast attosecond electron pulses for coherent excitation and scattering},
  author={Yalunin, Sergey V and Feist, Armin and Ropers, Claus},
  journal={Physical Review Research},
  volume={3},
  number={3},
  pages={L032036},
  year={2021},
  publisher={APS}
}

@article{morimoto2021coherent,
  title={Coherent scattering of an optically modulated electron beam by atoms},
  author={Morimoto, Yuya and Hommelhoff, Peter and Madsen, Lars Bojer},
  journal={Physical Review A},
  volume={103},
  number={4},
  pages={043110},
  year={2021},
  publisher={APS}
}

@article{ratzel2021controlling,
  title={Controlling quantum systems with modulated electron beams},
  author={R{\"a}tzel, Dennis and Hartley, Daniel and Schwartz, Osip and Haslinger, Philipp},
  journal={Physical review research},
  volume={3},
  number={2},
  pages={023247},
  year={2021},
  publisher={APS}
}

@article{zhang2025spontaneous,
  title={Spontaneous photon emission by shaped quantum electron wavepackets and the QED origin of bunched electron beam superradiance},
  author={Zhang, Bin and Ianconescu, Reuven and Friedman, Aharon and Scheuer, Jacob and Tokman, Mikhail and Pan, Yiming and Gover, Avraham},
  journal={Reports on Progress in Physics},
  volume={88},
  number={1},
  pages={017601},
  year={2025},
  publisher={IOP Publishing}
}

@article{karnieli2021superradiance,
  title={Superradiance and subradiance due to quantum interference of entangled free electrons},
  author={Karnieli, Aviv and Rivera, Nicholas and Arie, Ady and Kaminer, Ido},
  journal={Physical Review Letters},
  volume={127},
  number={6},
  pages={060403},
  year={2021},
  publisher={APS}
}

@article{karnieli2023jaynes,
  title={Jaynes-Cummings interaction between low-energy free electrons and cavity photons},
  author={Karnieli, Aviv and Fan, Shanhui},
  journal={Science advances},
  volume={9},
  number={22},
  pages={eadh2425},
  year={2023},
  publisher={American Association for the Advancement of Science}
}

@article{haffner2005scalable,
  title={Scalable multiparticle entanglement of trapped ions},
  author={H{\"a}ffner, Hartmut and H{\"a}nsel, Wolfgang and Roos, CF and Benhelm, Jan and Chek-al-Kar, D and Chwalla, M and K{\"o}rber, T and Rapol, UD and Riebe, M and Schmidt, PO and others},
  journal={Nature},
  volume={438},
  number={7068},
  pages={643--646},
  year={2005},
  publisher={Nature Publishing Group UK London}
}

@article{zang2016deterministic,
  title={Deterministic generation of large scale atomic W states},
  author={Zang, Xue-Ping and Yang, Ming and Ozaydin, Fatih and Song, Wei and Cao, Zhuo-Liang},
  journal={Optics express},
  volume={24},
  number={11},
  pages={12293--12300},
  year={2016},
  publisher={Optical Society of America}
}

@article{li2018dissipation,
  title={Dissipation-induced W state in a Rydberg-atom-cavity system},
  author={Li, Dong-Xiao and Shao, Xiao-Qiang and Wu, Jin-Hui and Yi, XX},
  journal={Optics Letters},
  volume={43},
  number={8},
  pages={1639--1642},
  year={2018},
  publisher={Optical Society of America}
}

@article{henke2025observation,
  title={Observation of quantum entanglement between free electrons and photons},
  author={Henke, Jan-Wilke and Jeng, Hao and Sivis, Murat and Ropers, Claus},
  journal={arXiv preprint arXiv:2504.13047},
  year={2025}
}

@article{preimesberger2025experimental,
  title={Experimental verification of electron-photon entanglement},
  author={Preimesberger, Alexander and Bogdanov, Sergei and Bicket, Isobel C and Rembold, Phila and Haslinger, Philipp},
  journal={arXiv preprint arXiv:2504.13163},
  year={2025}
}

@article{henke2025probing,
  title={Probing electron-photon entanglement using a quantum eraser},
  author={Henke, Jan-Wilke and Jeng, Hao and Ropers, Claus},
  journal={Physical Review A},
  volume={111},
  number={1},
  pages={012610},
  year={2025},
  publisher={APS}
}

\end{document}